\documentclass[aps,prd,preprint,groupedaddress]{revtex4-2}

\usepackage{amssymb,amsthm}
\usepackage{amsmath,amssymb,amsfonts,bm,amscd}
\usepackage{graphicx}
\usepackage{caption, subcaption}
\usepackage{multirow}
\usepackage{tensor}
\usepackage{tikz}

\usepackage{mathtools}
\usepackage{hyperref}

\newtheorem{conj}{Conjecture}
\newtheorem{cor}{Corollary}

\def\b{\beta}

\def\m{\mu}

\def\fkf{\mathfrak{f}}

\def\fsu{\mathfrak{su}}

\def\CI{\mathcal{I}}
\def\CM{\mathcal{M}}
\def\CN{\mathcal{N}}

\def\CT{\mathcal{T}}

\def\bbC{\mathbb{C}}
\def\bbH{\mathbb{H}}
\def\bbP{\mathbb{P}}

\def\bbZ{\mathbb{Z}}
\def\bbV{\mathbb{V}}

\def\tr{\mathrm{tr}}

\def\CH{{\cal H}}
\def\CI{{\cal I}}

\def\CM{{\cal M}}
\def\CN{{\cal N}}

\def\CP{{\cal P}}

\def\tr{\mathop{\rm tr}}

\def\beq#1\eeq{\begin{align}#1\end{align}}

\def\fsu{\mathfrak{su}}

\def\bbC{\mathbb{C}}

\def\bbZ{\mathbb{Z}}

\def\bfp{\mathbf{p}}
\def\bfx{\mathbf{x}}

\def\ch{\chi}

\def\beq#1\eeq{\begin{align}#1\end{align}}

\theoremstyle{definition}

\begin{document}
\title[4d Mirror Symmetry]{Mirror symmetry for 4d $A_1$ class-$\mathcal{S}$ theories: modularity, defects and Coulomb branch}


\author{Yiwen Pan}
\affiliation{Department of Physics,  Sun Yat-Sen University, Guangzhou, Guangdong, China}

\author{Wenbin Yan}
\affiliation{Yau Mathematical Sciences center, Tsinghua University, Beijing, China}

%
%
%
%
%


\begin{abstract}
This is the companion paper of the letter arXiv:2410.15695, containing all the details and series of examples on a 4d mirror symmetry for the class-$\mathcal{S}$ theories which relates the representation theory of the chiral quantization of the Higgs branch and the geometry of the Coulomb branch. We study the representation theory by using the 4d/VOA correspondence, (defect) Schur indices and (flavor) modular differential equations, and match the data with the fixed manifolds of the Hitchin moduli spaces. This correspondence extends the connection between Higgs and Coulomb branch of Argyres-Douglas theories, and can provide systematic guidance for the study of the representation theory of vertex operator algebras by exploiting  results from Hitchin systems.
\end{abstract}

\maketitle

\section{Introduction and main results}

The 3d mirror symmetry \cite{Intriligator:1996ex} is a remarkable duality between a pair of 3d  theories with 8 supercharges such that the Coulomb branch of one is the Higgs branch of the other one and vice versa. It is an invaluable tool to study moduli spaces of 3d theories, and also deeply related to 3d symplectic duality \cite{braden2012quantizations,braden2014quantizations} in mathematics. One may also ask whether similar mirror symmetry exists for 4d $\CN=2$ theories with also 8 supercharges. At first glance this seems impossible, as the Higgs branches and Coulomb branches in 4d have different geometries, and can never exchange roles like in 3d. However, inspired by the 3d symplectic duality where the quantization of 3d Higgs branch is connected to the cohomology of 3d Coulomb branch, one may view the 4d mirror symmetry as the relation between certain ``quantization'' of Higgs branch and the geometry of Coulomb branch of a given 4d $\CN=2$ theory. A natural choice is the vertex operator algebra (VOA) from the 4d/VOA correspondence \cite{Beem:2013sza} which is viewed as the chiral quantization of the Higgs branch \cite{Beem:2017ooy, Song:2017oew,Arakawa:2018egx}. 

Such 4d mirror symmetry was proposed and proved for 4d $\CN=2$ generalized Argyres-Douglas (AD) theories \cite{Shan:2023xtw,Shan:2024yas} (see also \cite{Fredrickson:2017yka, Fredrickson:2017jcf}). It is shown that given an AD theory $\CT$  there is a bijection between simple modules of $V[\CT]$ and $\bbC^\ast$ fixed points of $\CM_{C}[\CT]$. Furthermore, the space spanned by characters of simple modules is isomorphic to the cohomology of fixed points under the action of the modular group $SL(2,\bbZ)$. In certain cases, one can further show that the Zhu's $C_2$ algebra of $V[\CT]$ is isomorphic to the cohomology ring of $\CM_C[\CT]$. These results imply that there is indeed a deep relation between Higgs and Coulomb branches of 4d $\CN=2$ superconformal field theories (SCFTs).

In the letter \cite{Pan:2024hcz} we provided strong evidences for the correspondence between the chiral quantization of Higgs branch and geometry of Coulomb branch for a large class of 4d $\CN=2$ SCFTs: the $A_1$ class-$\mathcal{S}$ theories $\mathcal{T}_{g, n}$ constructed by compactification of 6d $(2,0)$ SCFTs on a genuse $g$ Riemann surface $\Sigma_{g,n}$  with $n$ regular punctures \cite{Gaiotto:2009we,Gaiotto:2009hg}. In this manuscript we give extensive details and further evidences for this relation.

The Higgs branch of $\CT_{g,n}$ is the Moore-Tachikawa variety \cite{Moore:2011}. The chiral quantization $V_{g,n}$ of the Higgs branch of $\CT_{g,n}$  has been constructed in \cite{Arakawa:2018egx}. It is expected that $V_{g,n}$ has a finite number of simple modules. Characters of theses simple modules,  which are meromorphic functions of the flavor fugacities and modular parameter $q$, correspond to (defect) Schur indices of the 4d theory, and satisfy a set of partial differential equations called the flavor modular differential equations (FMDEs) \cite{Zheng:2022zkm}. The quasi-modularity \cite{Pan:2023jjw} of FMDEs implies that these  characters belong to a representation $\bbV^\text{mod}_{g,n}$ of the modular group. Usually the dimension of $\bbV^\text{mod}_{g,n}$ is larger than the number of simple modules, where the extra solutions may be identified as characters of logarithmic modules. Ordinary modules are special simple modules whose characters are still convergent when setting all flavor fugacities to $1$. Such unflavored characters of ordinary modules satisfy a modular differential equation (MDE) \cite{Arakawa:2016hkg,Beem:2017ooy} and belong to a smaller representation $\bbV^\text{ord}_{g,n}$ of the modular group. Practically, we often use the correspondence between the 4d defects and VOA modules \cite{Beem:2013sza, Cordova:2017mhb,Cordova:2016uwk,Pan:2021mrw,Guo:2023mkn} to obtain characters of simple modules of $V_{g,n}$ \cite{Pan:2021mrw,Guo:2023mkn,Pan:2024bne}, and extract data of highest weight state from characters. The basis of $\bbV^\text{mod}_{g,n}$ or $\bbV^\text{ord}_{g,n}$ can be constructed using the Schur index and modular transformations \cite{Pan:2024bne}.

The geometry of the Coulomb branch of $\CT_{g,n}$ is described by the moduli space $\CM_{g,n}$ of the $SU(2)/\bbZ_2$ Hitchin system on $\Sigma_{g,n}$ \cite{Gaiotto:2009hg}. We are interested in the set of fixed manifolds  denoted by $\CM_{g,n}^T$ of $\CM_{g,n}$ under the $U(1)_r$ from the superconformal algebra. $\CM_{g,n}^T$'s  are studied in \cite{hitchin1987self} for $n=0$  and in \cite{boden1996moduli,nasatyr1995orbifold} for $n>0$. They provide a description for each fixed manifold and the critical value of the moment map of $U(1)_r$ on each fixed manifold, which will be crucial in matching with data from $V_{g,n}$. Notice that $\CM_{g,n}$ is also homeomorphic to the character variety $C_{g,n}$ on $\Sigma_{g,n}$ whose geometries are studied in \cite{hausel2008mixed}. We also find that information on fixed manifolds of $\CM_{g,n}$ can also be extracted from geometric data of $C_{g,n}$ which was not noticed by math community before.

Using the above data from both Higgs branch and Coulomb branch, we observe a bijection between the simple modules of $V_{g,n}$ and fixed manifolds of $\CM_{g,n}$ of the class-$\mathcal{S}$ theory $\CT_{g,n}$. The highest weight state of each simple module is completely determined by the value of moment map of the corresponding fixed manifold. Moreover, Jordan type of the modular matrices on $\bbV^\text{mod}_{g,n}$ are determined by dimensions of each fixed manifold and $g$. The later observation also provides an interpretation of logarithmic modules as certain elements in the cohomology of $\CM_{g,n}$. Our results show that there is a deep connection between the chiral quantization of the Higgs branch and geometry of Coulomb branch for a large class of 4d $\CN=2$ theories, suggesting a correspondence between representation theory of VOAs and geometry of Hitchin systems in more general terms. 

This paper is organized as the following. Section \ref{sec:modularmodules} discusses modular invariant modules of $V_{g,n}$ and how to obtain them. Section \ref{sec:hitchin} and \ref{sec:charvar} collects results on fixed manifolds of Hitchin moduli space $\CM_{g,n}$ and mixed Hodge polynomials of the character variety $C_{g,n}$ which will be used to match with module data. With all the preparation, we finally summarize our statements on the mirror symmetry in section \ref{sec:results}.  We then provide examples of our conjectures in section \ref{sec:examples}. Generalizations to higher rank class-$\mathcal{S}$ theories are discussed in \ref{sec:higherrank}. Finally, generalizations and future directions will be discussed in \ref{sec:outlook}.

\section{Modular invariant simple modules of $V_{g,n}$}
\label{sec:modularmodules}

In this section we review modular invariant simple modules of $V_{g,n}$ and how to obtain their characters from FMDE or (defect) Schur indices. By modular invariant simple modules we mean simple modules whose characters belong to a finite dimensional representation of the modular group of $V_{g,n}$, and we will call them simple modules for short in these paper. 

\subsection{Characters of simple modules of $V_{g,n}$ from Schur indices}

Any 4d $\mathcal{N} = 2$ SCFT $\mathcal{T}$ contains a subsector of protected local operators (Schur operators) which can be endowed with the structure of a 2d VOA $V[\mathcal{T}]$ with central charge $c=-12 c_\text{4}$ \cite{Beem:2013sza}. Here $c_\text{4d}$ is the $c$ central charge of the 4d theory $\CT$.
When $\mathcal{T}$ enjoys a  continuous flavor symmetry $\mathfrak{f}$, $V[\mathcal{T}]$ contains an affine subalgebra $\widehat{\mathfrak{f}}_{k}$ at level $k=- \frac{1}{2} k_\text{4d}$ where $k_\text{4d}$ is the 4d anomaly coefficient of $\fkf$. Through this correspondence, many physical observables in the 4d theory $\mathcal{T}$ are mapped to the representation data of the associated VOA $V[\mathcal{T}]$. One such example is the Schur index $\mathcal{I}[\mathcal{T}]$ \cite{Gadde:2011ik,Gadde:2011uv},
\begin{equation}
	\mathcal{I}[\mathcal{T}] =  \operatorname{tr}_{\mathrm{Schur}} (-1)^F q^{E - R}b^{h_\fkf}	,
\end{equation}
which is identified with the (normalized) vacuum character $q^{c/24}\ch_0$ of $V[\mathcal{T}]$. Here $h_\fkf$ collectively denotes the Cartan generators of $\fkf$ and the flavor fugacities $b$ collectively represents flavor fugacities $b_i$. Similarly, Schur indices with defects inserted correspond to linear combinations of characters of simple modules \cite{Cordova:2017mhb}. Therefore, the 4d/VOA correspondence is a powerful tool to study the representation theory of $V[\CT]$ when physical observables like (defect) Schur indices can be computed explicitly.
For all indices (characters), one can take the unflavored limit by sending all $b_i$'s to $1$, however, the result may not always be convergent.


In particular, for $A_1$ class-$\mathcal{S}$ theory $\CT_{g,n}$,
the flavored Schur index $\mathcal{I}_{g,n} = \ch_0[V_{g,n}]$ has a closed-form expression in terms of Eisenstein series $E_k\big[\substack{\pm 1 \\ b}\big]$ \cite{Pan:2021mrw}, 
\begin{align}
	\mathcal{I}_{g,n > 0} = & \ \frac{i^n}{2} \frac{\eta(\tau)^{2g - 2 + n}}{\prod_{j = 1}^{n}\vartheta_1(2 \mathfrak{b}_j)} \sum_{\alpha_j = \pm 1} \left(\prod_{j = 1}^{n} \alpha_j\right)
	\sum_{k = 1}^{2g - 2 + n}\lambda_k^{(2g - 2 + n)} E_k \begin{bmatrix}
    (-1)^n \\ \prod_{j = 1}^{n} b_j^{\alpha_j}
	\end{bmatrix}\ ,\\
	\mathcal{I}_{g, 0} = & \ \frac{1}{2} \eta(\tau)^{2g - 2}\sum_{k = 1}^{g - 1} \lambda_{2k}^{(2g - 2)} \bigg(E_{2k} + \frac{B_{2k}}{(2k)!}\bigg) \ ,
\end{align}
where $\lambda$'s are some rational numbers defined by a set of recursion relations, and $b_i = e^{2\pi i \mathfrak{b}_i}$ denotes the flavor fugacity of the $SU(2)$ flavor symmetry associated to the $i$-th puncture on the corresponding Riemann surface $\Sigma_{g,n}$. Such closed form expression made the study of modular properties of characters possible.

A vortex surface defect of $\CT_{g,n}$ can be constructed by coupling the theory to a $\mathcal{T}_{0,3}$, then turning on a position-dependent vacuum expectation value (VEV) for a  $U(1)$ baryonic operator, and finally flowing the system to the infrared \cite{Gaiotto:2012xa}. 
Let $\kappa$ be the vorticity of the defect,
the defect Schur index $\mathcal{I}_{g,n}^\text{vortex}(\kappa)$ is certain residue of the Schur index $\mathcal{I}_{g,n + 1}$  \cite{Gaiotto:2012xa,Cordova:2017mhb,Alday:2013kda,Nishinaka:2018zwq},
\begin{equation}
	\mathcal{I}_{g,n}^\text{vortex}(\kappa) = 2(-1)^\kappa q^{- \frac{(\kappa + 1)^2}{2}}\mathop{\operatorname{Res}}_{b_{n + 1} \to q^{\frac{1 + \kappa}{2}}} \frac{\eta(\tau)^2}{b_{n + 1}}  \mathcal{I}_{g, n + 1}(b_1, \ldots, b_{n + 1}) \ .
\end{equation}
Closed-form of $\mathcal{I}_{g,n}^\text{vortex}(\kappa)$ is also known \cite{Pan:2021mrw},
\begin{align}\label{eq:vortex-defect-index}
  \mathcal{I}^{\text{vortex}}_{g, n > 0}(\kappa) = & \ (-1)^\kappa \frac{\eta(\tau)^{n + 2g - 2}}{\prod_{i = 1}^n\vartheta_1(2 \mathfrak{b}_i)}\\
  & \ \times \sum_{\alpha_i = \pm}\left(\prod_{i = 1}^{n}\alpha_i\right)
  \sum_{\ell = 1}^{n + 1 + 2g - 2}\tilde\lambda^{n + 1 + 2g - 2}_\ell(\kappa + 1) E_\ell \left[\begin{matrix}
      (-1)^{n + \kappa}\\ \prod_{i = 1}^{n}b_i^{\alpha_i}
  \end{matrix}\right]\ , \nonumber\\
	\mathcal{I}^\text{vortex}_{g,0}(\kappa) = & \ \eta(\tau)^{2g - 2} \sum_{\ell = 0}^{g - 1} c_\ell(\kappa) E_{2\ell} \begin{bmatrix}
		(-1)^\kappa \\ 1
	\end{bmatrix}  \ ,
\end{align}
for some rational numbers $\tilde \lambda_\ell^{(2g - 2 + n + 1)}$ and $c$. For our purpose, we focus on the defects with even vorticity $\kappa$. When $\kappa = 0$, the defect is effectively absent, whose index recovers the original Schur index $\mathcal{I}_{g, n}$.

Since all duality frames of $A_1$ class-$\mathcal{S}$ theory $\CT_{g,n}$ have Lagrangian description with $3g-3 + n$ $SU(2)$ gauge groups coupling $2g - 2 + n$ copies of $\mathcal{T}_{0,3}$, one can insert Wilson lines charged under the various $SU(2)$ gauge groups into the index. When the Wilson line is charged under only one $SU(2)$ gauge group, there are two possible types: type II if $\CT_{g,n}$ becomes two decoupled $A_1$ class-$\mathcal{S}$ theories in the zero coupling limit of this $SU(2)$ gauge group, and type I otherwise. The  index with a type I Wilson line in the spin-$j$ representation inserted is \cite{Guo:2023mkn},
\begin{align}\label{Wilson-index-1-general}
  \langle W_{j \in \mathbb{Z}}\rangle^\text{(I)}_{g \ge 1, n}
  = & \ \mathcal{I}_{g,n}
  - \frac{1}{2}\left[
    \prod_{i = 1}^{n} \frac{i \eta(\tau)}{\vartheta_1(2 \mathfrak{b}_i)}
  \right]
    \sum_{\substack{m = - j\\m \ne 0}}^{+ j}
    \left[\frac{\eta(\tau)}{q^{m/2} - q^{-m /2}}\right]^{2g - 2}
    \prod_{i = 1}^{n} \frac{b_i^m - b_i^{-m}}{q^{m/2} - q^{- m /2}} \ ,\\
  \langle W_{j \in \mathbb{Z} + \frac{1}{2}}\rangle^\text{(I)}_{g \ge 1, n} = & \ 0 \ .
\end{align}
For a type II Wilson line, if the two decoupled theories are $\CT_{g_1, n_1}$ and $\CT_{g_1,n_1}$, then
\begin{align}
  & \ \langle W_j\rangle_{g_1, n_1; g_2, n_2}^{\text{(II)}} \nonumber \\
  = & \ \mathcal{I}_{g_1 + g_2, n_1 + n_2 - 2}\delta_{j \in \mathbb{Z}} \nonumber\\
  & \ + \frac{\eta(\tau)^{2(g_1 + g_2) + (n_1 + n_2 - 2) - 2}}{
    2\prod_{i = 1}^{n_1 + n_2 - 2}\vartheta_1(2 \mathfrak{b}_i)
  }\\
  & \ \qquad \times \sum_{\substack{m = - j \\ m\ne 0}}^j \sum_{\vec \alpha} \left(\prod_{i=1}^{n_1 + n_2 - 2}\alpha_i\right)
  \sum_{\ell = 0}^{\operatorname{max}(n_i + 2g_i - 2)}
    \Lambda_\ell^{(g_1, n_1; g_2, n_2)}(\mathbf{b}^{2m}, q^{2m})
    E_\ell \begin{bmatrix}
    1 \\ \prod_{i}^{n_1 + n_2 -2} b_i^{\alpha_i}
  \end{bmatrix} \ , \nonumber
\end{align}
where $\Lambda$ denote some rational functions of $q, b_i$.  Wilson line indices are linear combinations of characters with coefficients being rational functions of $q, b_i$.
All the above line and surface defect indices all have smooth $b_i \to 1$ limit.

\subsection{Modular invariance}

By construction, the VOA $V[\mathcal{T}]$ is quasi-lisse \cite{Arakawa:2016hkg}, whose modules share similar modularity properties as rational VOAs. A quasi-lisse VOA has various type of modules. The simplest ones are ordinary modules - simple modules whose weight spaces at any given conformal weight $h$ are finite dimensional. The vacuum module is one example of ordinary module. There are also simple but non-ordinary modules, whose weight spaces are infinite dimensional at certain $h$. Finally, modularity requires existence of logarithmic modules.


It is shown that the unflavored limit of the vacuum character (Schur index) $\ch_0$ together with characters of all ordinary modules of  $V[\CT]$ satisfy  an MDE \cite{Arakawa:2016hkg,Beem:2017ooy} \footnote{There may be multiple MDEs annihilating $\ch_0$. We only concern ourselves with the one of the lowest order.},
\begin{equation}
\label{eq:MDEofT}
  D_q^{(N)}\ch_0 + \sum_{r = 0}^{N - 1} \phi_{N - r}(\tau) D_q^{(r)}\ch_0 = 0 \ ,
\end{equation}
where $D_q^{(k)} \coloneqq \partial_{2k - 2}\partial_{2k -4} \cdots \partial_{(0)}$ with $\partial_{(k)} \coloneqq q\partial_q + k E_2(\tau)$, and the coefficient $\phi_{n}(\tau)$ denotes a modular form \footnote{As a meromorphic function of the complex structure moduli $\tau$, $\phi_n(\tau)$ may contain poles.} of weight $2n$ with respect to the modular group $\Gamma$ of $V[\mathcal{T}]$. Precise form of $\phi_n$ varies according to the theory $\CT$.
Under the action of the modular group $\Gamma$, the equation transforms covariantly, therefore all its solutions span a representation $\bbV^\text{ord}[\mathcal{T}]$ of $\Gamma$ of dimension $N$. 
Usually equation \ref{eq:MDEofT} contains also solutions with $\log q$ terms besides unflavored characters of ordinary modules. These logarithmic solutions could be viewed as characters of certain logarithmic modules. Unlike rational VOA, characters of ordinary modules alone are not modular invariant. The modular invariance is only achieved with the inclusion of characters of logarithmic modules.

The full vacuum character (Schur index), together with characters of other simple modules of $V[\CT]$ satisfy FMDEs which is a set of partial differential equations with quasi-Jacobi forms as coefficients. Schematically, an FMDE looks like
\begin{equation}
	\sum_{\substack{k, \ell_1, ..., \ell_r \\ 2k + \ell_1 + ... + \ell_r \le 2n}} \phi_{k, \ell_1, ..., \ell_r}(q, b_i) D_{b_1}^{\ell_1} ... D_{b_r}^{\ell_r} D_q^{(k)} \ch_0 = 0 \ ,
\end{equation}
where $b_i$'s are flavor fugacities $b_i$ associated to Cartans of flavor symmetry of $\CT$,  $D_{b_i} \coloneqq b_i \partial_{b_i}$, and $\phi_{k, \ell_1, \cdots, \ell_r}$ is a quasi-Jacobi form. Under the modular group, $D_{b_i}$ is of modular weight $1$, $D_q^{(k)}$ of $2k$, and the weight of $\phi_{k, \ell_1, \ldots, \ell_r}(q, b_i)$ is determined by requiring homogeneity of the whole equation. These equations again come from suitable null states in VOA $V[\mathcal{T}]$ which is equivalent to relations among Schur operators. Similar to the unflavored case, we expect that the solutions of the FMDEs also form a representation $\bbV[\CT]$ of the modular group $\Gamma$ \footnote{The modular group $\Gamma$ in the flavored case may be different from the one relevant in the unflavored case of the same theory: in general, flavor refinement results in a subgroup of the latter}. However, this is a highly non-trivial statement, since an FMDE is in general not covariant under the modular group $\Gamma$; instead, it exhibits ``quasi-modular property'' where an FMDE $\text{eq}^{(w)}$ of weight $w$ transforms into a sum of FMDEs of equal or lower weights \cite{Zheng:2022zkm},
\begin{equation}
	\text{eq}^{(w)} \to \tau^w \text{eq}^{(w)} + \sum_{i}\tau^{w - 1} \mathfrak{b}_i \text{eq}^{(w-1)}_i
	+ \sum_{i,j} \tau^{w - 2}\mathfrak{b}_i \mathfrak{b}_j \text{eq}^{(w-2)}_{ij} + \cdots \ .
\end{equation}
Note that $\tau = \frac{1}{2\pi i}\log q$, $\mathfrak{b}_i = \frac{1}{2\pi i}\log b_i$.

Quasi-modularity then implies that a solution to $\text{eq}^{(w)}$ should simultaneously solve all the $\text{eq}^{(w')}_{ij\cdots}$ that appear on the right. The additional FMDEs are likely to arise from additional null states, and the modular transformation of the FMDEs $\text{eq}^{(w)}, \text{eq}^{(w')}_{ij\cdots}$ can be viewed as an action of $\Gamma$ on these null states.

FMDEs also have two types of solutions. Solutions without terms proportional to $(\log q)^n$ and $(\log b_i)^n$ are (linear combinations) of characters of simple modules, while logarithmic solutions are also present because for invariance. A non-logarithmic solution can be expanded as a $q$-series $a_0(b_i)q^{-h+c/24}+\cdots$, from the leading term of which one can read off both the conformal weight $h$ and flavor weight/Dynkin label of the highest weight state of the corresponding simple modules.

For $\CT_{g,n}$ with small values of $g, n$, the corresponding null states and/or FMDEs can be worked out explicitly, then use them to construct the representation $\mathbb{V}_{g,n}^\text{mod}$ of $\Gamma$\footnote{$\Gamma$ is $SL(2,\bbZ)$ when $n$ is even, is $\Gamma^0(2)\subset SL(2,\bbZ)$ when $n$ odd.}. However, as $g$ or $n$ get larger, it is increasingly difficult to find out the precise form of all the FMDEs. In that case, one can build a basis for a subspace $\mathbb{V}_{g,n}^\text{ord}$ of $\bbV_{g,n}^\text{mod}$ starting from the closed form expressions of characters of simple ordinary modules from defected Schur indices mentioned in the previous subsection, then apply modular transformation on the indices (characters) to complete the whole basis. In the end, one finds the following basis of $\mathbb{V}_{g,n}^\text{ord}$ 
\begin{equation}
	\begin{aligned} \left\{\mathbf{T}_{(2)}^{\ell}S_{(2)}\mathcal{I}_{g,n}\right\}_{\ell=0}^{g}& \ \cup\left\{\mathbf{T}_{(2)}^{\ell}S_{(2)}\mathbf{T}_{(2)}^{g}S_{(2)}\mathcal{I}_{g,n}\right\}_{\ell=0}^{g+2}\cup\left\{\mathbf{T}_{(2)}^{\ell}S_{(2)}\mathbf{T}_{(2)}^{g+2}S_{(2)}\mathbf{T}_{(2)}^{g+1}S_{(2)}\mathcal{I}_{g,n}\right\}_{\ell=0}^{g+4}\\ &\cup\cdots  \cdots\cup\left\{\mathbf{T}_{(2)}^{\ell}S_{(2)}\mathbf{T}_{(2)}^{3g-3+n-2}\cdots S_{(2)}\mathbf{T}_{(2)}^{g+1}S_{(2)}\mathcal{I}_{g,n}\right\}_{\ell=0}^{3g-3+n}.\end{aligned}
\end{equation}
Here $\mathbf{T}_{(2)}\equiv e^{-\frac{\pi i}{3}(g-n-1)} T^2 -\mathbf{1}$, $S_{(2)} \coloneqq STS$, which are always elements of both $SL(2,\bbZ)$ or $\Gamma^0(2)\subset SL(2,\bbZ)$. The representation matrices for $S$ and $T$ (resp. $STS$ and $T^2$) can then be computed explicitly.

\section{Fixed manifolds of moduli spaces of Higgs bundles}
\label{sec:hitchin}

In this section we review some information on Higgs bundles and their fixed points under the $U(1)$($\bbC^\ast$) action. We will mainly focus on rank-one case. Higher rank cases are discussed in \cite{MR2308696}.

Following \cite{Gaiotto:2009hg}, the Coulomb branch of $\CT_{g,n}$ is described by the moduli space $\CM_{g,n}$ of $SU(2)/\bbZ_2$ Hitchin system on $\Sigma_{g,n}$. Specifically $\CM_{g,n}$ is the solution space of the Hitchin equations on $\Sigma_{g,n}$ \cite{hitchin1987self}
\begin{align}
&F_A+[\varphi,\varphi^{\dag}]=0,\\
&\bar{\partial}_{A}\varphi=0,
\end{align}
where $F_A$ is the curvature of the gauge field $A=A_zdz+A_{\bar{z}}d\bar{z}$ and $\varphi$ is the adjoint scalar. At the $i$-th regular puncture, the pair $(A,\varphi)$ is singular with the following asymptotic form around the puncture
\begin{align}
A&\sim \alpha_i d\theta, \\
\varphi& \sim \frac{f}{z} + \mathrm{regular}.
\end{align}
Here $f$ is a nilpotent element in $\fsu(2)$ Lie algebra.
In math literature, $\CM_{g,n}$ is also called the moduli space of Higgs bundles, and $\alpha_i$ is the parameter specifies the parabolic structure of the Higgs bundle around the $i$-th puncture.

The $U(1)_r$ R- symmetry acts non trivially on $\CM_{g,n}$ which simply rotate the Higgs field $\varphi$ by a phase. 
Given a pair $(A,\varphi)\in \CM_{g,n}$, the moment map $\mu$ of the $U(1)_r$ is defined as
\begin{equation}
\mu=\frac{i}{2\pi}\int_{\Sigma_{g,n}}\tr(\varphi \varphi^\dag)dzd\bar{z},
\end{equation}
which is a Morse function on $\CM_{g,n}$. Therefore
the set of the $U(1)_r$ fixed manifolds $\CM^T_{g,n}$ and critical values of $mu$ on each fixed manifolds are important in studying the topology of $\CM_{g,n}$, and we summarize results here.

Firstly we consider cases with $n = 0$ punctures, the moduli space of Higgs bundles is non-trivial when $g\geq2$. There are exactly $g$ fixed manifolds as described by proposition 7.1 of \cite{hitchin1987self}.
\begin{itemize}
\item The fixed manifold $M_0$ with moment map $\mu=0$ is diffeomorphic to the moduli spaced of stable rank-2 bundles of odd degree and fixed determinant over $\Sigma_{g,0}$.
\item There are $g-1$ fixed manifolds $\{M_d\}_{1\leq d\leq g-1}$ with $\mu = d-\frac{1}{2}$ and index $\lambda_d=2(g+2d-2)$. $\CM_d$ is diffeomorphic $\tilde{S}\Sigma_{g,0}$, a $2^{2g}-$ fold covering of $S^{2g-2d-1}\Sigma_{g,0}$ where $S^d M$ is the $d$-th symmetric product of $M$.
\end{itemize}
Note that moment map formula $\mu(M_{d})$ features a $\frac{1}{2}$ jump at $d = 0$. The maximal value of $\mu$ is $\mu_\text{max} = g - \frac{3}{2}$, and the minimal value $\mu_\text{min} = 0$.

Now we add $n > 0$ regular singularities to the Riemann surface. In particular, we consider the cases where $2g - 2 + n > 1$, and the $\mathcal{N} = 4$ $SU(2)$ theory, corresponding to the $(g, n) = (1,1)$ case but with the free hypermultiplet decoupled. For each puncture, we need  parameter $0<\alpha_i<\frac{1}{2}$  to describe the parabolic structure on the $i-$th puncture. Fixed manifolds are classified in \cite{boden1996moduli,nasatyr1995orbifold}. Following the notation of \cite{boden1996moduli}, define $e_i\in \{0,1\}$ for the $i-$th puncture and $\beta_i(e_i, \alpha_i) = e_i+(-1)^{e_i}\alpha_i$. Write $e=(e_1,\cdots,e_n)$, $\alpha=(\alpha_1,\cdots,\alpha_n)$ and $\beta=(\beta_1,\cdots,\beta_n)$. Also define $|e|=\sum_{i=1}^n e_i$. Then the fixed manifolds are given as follows.
\begin{itemize}
\item There is a set of fixed manifolds $M_{d,e}$ where $d$ is an integer, such that the sequence $e$ and $d$ should satisfy the constraint
\begin{equation}\label{eq:constraint}
	-\sum_{i=1}^n\beta_i(e_i, (n + 1)^{-i})<d\leq g-1-|e|/2.
\end{equation}
On $M_{d,e}$ the moment map $\mu = \mu_{d,e}\coloneqq d+\sum_{i=1}^n\beta_i$. The index $\lambda_{d,e}=2(n+2d+g-1+|e|)$. $M_{d,e}$ is diffeomorphic to $\tilde{S}^{h_{d,e}}\Sigma_{g,n}$, a $2^{2g}$ cover of the $S^{h_{d,e}}\Sigma_{g,n}$ with $h_{d,e}=2g-2-2d-|e|$. Here $S^{h_{d,e}} \Sigma_{g,n}$ denotes the $h_{d,e}$-fold symmetric product of $\Sigma_{g,n}$. 
\item When $g\geq1$, there is an additional fixed point $M_0$ with $\mu=0$ which is diffeomorphic to the moduli space of stable rank-2 bundles  over $\Sigma_{g,n}$, with dimension $\dim M_0 = 3g - 3 + n$.
\end{itemize}
A few remarks follow. In the above definition, the fixed manifolds $M_{d,e}$ does not depend on the parabolic structure parameters $\{\alpha_i\}$. However, the values of the moment map $\mu$ on each fixed manifold does depend on $\alpha_i$, which will be mapped to the VOA data. The constraint (\ref{eq:constraint}) implies
\begin{equation}
	\mu_{d,e} > 0, \qquad h_{d,e} \ge 0 \ .
\end{equation}
Hence, for $g \ge 1$, the moment map assumes its minimum on $M_0$ with $\mu_\text{min} = 0$, while for $g = 0$, $\mu_\text{min} = \alpha_1 - \prod_{i = 2}^{n}\alpha_i$ on $M_{-(n - 1), (0,1, \cdots, 1)}$ where $\lambda_{d,e} = 0$.

For any $g, n$, consider $e = (1, \cdots 1)$ and hence $|e| = n$. Then we have
\begin{align}
	& \ \sum_{i = 1}^{n}\beta_i(e_i, (n + 1)^{-i})
	= n - \frac{(n + 1)^{-(n + 1)}((n + 1)^(n + 1) - 1)}{n}\\
	\Rightarrow & \ \bigg\lceil - \sum_{i = 1}^{n}\beta_i(e_i, (n + 1)^{-i}) \bigg\rceil = 1 - n \ .
\end{align}
As a result, there are always solutions of integer $d$ to the constraint (\ref{eq:constraint}) for this special $e = (1, \cdots, 1)$,
\begin{equation}
	1 - n \le d \le g - 1 - \frac{n}{2} \ .
\end{equation}
Therefore, there always exist the fixed manifolds $M_{d, (1,\cdots, 1)}$ where $1 - n \le d \le g - 1 - \frac{n}{2}$. In particular, when $n$ is positive even and $d = g - 1 - \frac{n}{2}$,
\begin{equation}
	\mu_{d = g - 1 - \frac{n}{2}, (1, \cdots, 1)} = g - 1 + \frac{n}{2} - \sum_{i = 1}^{n}\alpha_i
	= \max_{e} \mu_{d = g - 1 - \frac{n}{2}, e} \ .
\end{equation}
Note that at a different $d$ there may be some $\mu_{d,e} \ge \mu_{d = g - 1 - \frac{n}{2}, (1, \cdots, 1)}$. When the parabolic parameters further satisfy $\alpha_n < \cdots <\alpha_2 < \alpha_1 < \frac{1}{4}$, $\mu_{g - 1 - \frac{n}{2}, (1, \cdots, 1)} = \max_{e, d}\mu_{d,e}$ . Therefore we denote $\mu_\text{max} \coloneqq \mu_{g - 1 - \frac{n}{2}, (1, \cdots, 1)}$ for positive and even $n$. Hence, for other $d, e$,
\begin{equation}
	\mu_{d,e} - \mu_\text{max} = d + |e| - (g - 1 + \frac{n}{2}) + \sum_{i = 1}^{n}[(-1)^{e_i} + 1]\alpha_i \ ,
\end{equation}
where the coefficients of $\alpha_i$ are all non-negative. In particular,
\begin{equation}
	\mu_{d, (1, \cdots, 1)} = d + n - (g - 1 + \frac{n}{2}) \in \{2 - g - \frac{n}{2}, \cdots, -2, -1, 0\}
\end{equation}

Similarly, for the theories we consider, there are always the fixed manifolds
\begin{equation}
	M_{d, (0, 1, \cdots, 1)}, \qquad 1 - n \le d \le g - \frac{1}{2} - \frac{n}{2} \ .
\end{equation}
In particular, when $n$ is positive odd and $d = g - \frac{1}{2} - \frac{n}{2} \in \mathbb{Z}$,
\begin{equation}
	\mu_{g - \frac{1}{2} - \frac{n}{2}, (0, 1, \cdots, 1)}
	= g - \frac{3}{2} + \frac{n}{2} + \alpha_1 - \sum_{i = 2}^{n}\alpha_i
	= \max_{e} \mu_{g - \frac{1}{2} - \frac{n}{2}, e} \ .
\end{equation}
We then denote $\mu_\text{max} = \mu_{g - \frac{1}{2} - \frac{n}{2}, (0, 1, \cdots, 1)}$. When the parabolic parameters further satisfy $\alpha_n < \cdots <\alpha_2 < \alpha_1 < \frac{1}{4}$, $\mu_\text{max} = \max_{e, d}\mu_{d,e}$. The difference
\begin{equation}
	\mu_{d, e} - \mu_\text{max} = d + |e| - (g - \frac{3}{2} + \frac{n}{2}) + [(-1)^{e_1} - 1]\alpha_1 + \sum_{i = 2}^{n}[(-1)^{e_i} + 1]\alpha_i \ .
\end{equation}
The coefficients of $\alpha_{i = 2, \ldots, n}$ are all non-negative, while the coefficient of $\alpha_1$ is non-positive. In particular,
\begin{equation}
	\mu_{d, (0, 1,\ldots, 1)} - \mu_\text{max} 
	= d - (g - \frac{1}{2} - \frac{n}{2})
	\in \{\frac{3}{2}- \frac{n}{2} - g, \ldots, -2, -1, 0\}
\end{equation}


\section{Mixed Hodge polynomials of character varieties}
\label{sec:charvar}

To complete the last piece of the correspondence, we review some known results on the mixed Hodge polynomials of the $SU(2)/\bbZ_2$ character varieties. Fix a tuple of $n$ partitions $\bfp=(p^1,p^2,\cdots,p^n)$ of size $k$. Let $G$ be $U(k)$ or $SU(k)/\bbZ_k$. A $G$ character variety $C_{\bfp}^{G}$ of the Riemann surface $\Sigma_{g,n}$ is
\begin{align}
C_{\bfp}^G:=\{A_1&,B_1,\cdots,A_g,B_g\in G,X_1\in C_1,\cdots,X_n\in C_n| \\
&(A_1,B_1)\cdots(A_g,B_g)X_1\cdots X_n=I_k\} / G.
\end{align}
Here $C_i\subset G$ is the semisimple conjugacy class fixed by the partition $p^i$. $(A,B):=ABA^{-1}B^{-1}$ and $I_k$ is the identity matrix.

The (compactly supported) \emph{mixed Hodge polynomials} of $C_{\bfp}^G$ is
\begin{equation}
H_c(C^G_{\bfp};x,y,t):=\sum h^{i,j;k}_c(C^G_{\bfp})x^iy^jt^k,
\end{equation}
where $h^{i,j;k}_c(C_\bfp^G)$ are the compactly supported mixed Hodge numbers \cite{deligne1971theorie, deligne1978theorie}. It is a deformation of the compactly supported Poincare polynomial $P_c(C_\bfp^G;t)=H_c(C_\bfp;1,1,t)$.  The \emph{pure part} of $H_c$ is the polynomial
\begin{equation}
PH_c(C_\bfp^G ; x,y):=\sum h^{i,j;i+j}_c(C_\bfp^G)x^iy^j.
\end{equation}
Because $h^{i,j;k}_c(C_\bfp^G)=0$ unless $i=j$, $H_c(C^G_{\bfp};x,y,t)$ is a polynomial in variables $xy$ and $t$, and one can write
\begin{equation}
H_c(C^G_{\bfp};q,t):=H_c(C^G_{\bfp};x=\sqrt{q},y=\sqrt{q},t),
\end{equation}
and
\begin{equation}
PH_c(C^G_{\bfp};q):=PH_c(C^G_{\bfp};x=\sqrt{q},y=\sqrt{q})
\end{equation}
for simplicity.

$H_c(C^G_{\bfp};q,t)$ and $PH_c(C^G_{\bfp};q)$ can be computed using the  Macdonald symmetric function \cite{Hausel_2011}, which we will recall. Let the Cauchy function of $\Sigma_{g,n}$ be
\begin{equation}
\label{eq:Cauchy}
\Omega(z,w):=\sum_{p\in \CP}\CH_p(z,w)\prod_{i=1}^n \tilde{H}_p(\bfx_i;z^2,w^2),
\end{equation}
where $\CP$ is the set of all partitions, $\tilde{H}_p(\bfx;z^2,w^2)$ is the modified Macdonald symmetric function \cite{garsia1996remarkable}, and $\CH_p(z,w)$ is the $z$, $w$ deformation of the $(2g-2)$-th power of the hook polynomial
\begin{equation}
\CH_p(z,w):=\prod_{\square \in p} \frac{ (z^{2a(\square)+1}-w^{2l(\square)+1})^{2g} }{ (z^{2g+2}-w^{2l(\square)})(z^{2a(\square)}-w^{2l(\square)+2})) }.
\end{equation}
Here the product runs over all boxes $\square$ in $p$. $a(\square)$ and $l(\square)$ are arm- or leg-length of the corresponding box.

Given a tuple of $n$-partitions $\bfp=(p^1,\cdots,p^n)\in\CP^n$, define
\begin{equation}
\bbH_\bfp(z,w):=(z^2-1)(1-w^2)\langle \mathrm{Plog}[\Omega(z,w)],h_\bfp\rangle.
\end{equation}
Here $h_\bfp:=h_{p^1}(\bfx_1)\cdots h_{p^n}(\bfx_n)$ the product of complete symmetric functions, $\langle\cdot,\cdot\rangle$ is the extended Hall pairing
\begin{equation}
\langle a_1(\bfx_1)\cdots a_n(\bfx_n),b_1(\bfx_1)\cdots b_n(\bfx_n)\rangle
=\langle a_1,b_1\rangle\cdots \langle a_n,b_n\rangle,
\end{equation}
where $\langle a_i,b_i\rangle$ is the Hall pairing on symmetric functions over $\bfx_i$ \cite{macdonald1998symmetric}. $\mathrm{Plog}$ is the plythesitc logrithm
\begin{equation}
\mathrm{Plog}[f(x_1,x_2,\cdots)]=\sum_{n=1}^{\infty}\frac{\mu(n)}{n}\log f(x^n_1,x^n_2,\cdots),
\end{equation}
where $\mu$ is the Mobius function.

The most crucial observation of \cite{Hausel_2011} is that for $G=U(k)$, the mixed Hodge polynomial and its pure part of $C^{U(k)}_{\bfp}$ can be expressed in terms of $\bbH_\bfp$, more precisely
\begin{equation}
H_c(C^{U(k)}_{\bfp};q,t) = (t\sqrt{q})^{d_\mu} \bbH_{\mu}\left(-\frac{1}{\sqrt{q}},t\sqrt{q}\right),
\end{equation}
and
\begin{equation}
PH_c(C^{U(k)}_\bfp;q,t) = (q)^{d_\mu} \bbH_{\mu}\left(0,\sqrt{q}\right),
\end{equation}
where $d_\mu$ is the dimension of $C^{U(k)}_{\bfp}$. To get the mixed Hodge polynomial of $C^{SU(k)/\bbZ_k}_{\bfp}$, one simply removes a factor $(t\sqrt{q})^{2g}(1+qt)^{2g}$ from $C^{U(k)}_{\bfp}$ which corresponds to the mixed Hodge polynomial of $C^{U(1)}_{\bfp}$. By the definition of the pure part of mixed Hodge polynomials, $PH_c(C^{SU(k)/\bbZ_k}_{\bfp})$ differs from $PH_c(C^{U(k)}_{\bfp})$ only by a factor $q^{2g}$.

In this paper we consider only $k=2$, and each puncture has only one choice of partition $[1^2]$. The $\bbH$-polynomial for $g$ and $n>0$ is
\begin{equation}
\begin{split}
\bbH_{g,n}
=&\frac{(z^3-w)^{2g}(1+z^2)^n}{(z^2-w^2)(z^4-1)}+
\frac{(w^3-z)^{2g}(1+w^2)^n}{(w^2-z^2)(w^4-1)}\\
&+\frac{(z-w)^{2g}2^{n-1}}{(w^2-1)(z^2-1)}.
\end{split}
\end{equation}
To get $\bbH$-polynomial for $g$ and $n=0$, one can start with a genus $g$ Riemann surface with $1$ puncture with partition $[2]$, and the result is \cite{hausel2008mixed}
\begin{equation}
\label{eq:mhpg0}
\begin{split}
\bbH_{g,0}
=&\frac{(z^3-w)^{2g}}{(z^2-w^2)(z^4-1)}+
\frac{(z-w^3)^{2g}}{(w^2-z^2)(1-w^4)} \\
& -\frac{(z-w)^{2g}}{2(z^2-1)(1-w^2)}
-\frac{(z+w)^{2g}}{2(z^2+1)(1+w^2)}.
\end{split}
\end{equation}
Physically this means that one can remove a puncture by fully close it.
One can then write down $PH_c(C^{SU(2)/\bbZ_2}_{g,n};q,t)$ from $\bbH$-polynomials.

To better compare with fixed manifold, we need a renormalized version of $PH_c(C^{SU(2)/\bbZ_2}_{g,n};q)$
\begin{equation}
\label{eq:renormalizedMHP}
\begin{split}
P_{g,n}(q)&:=q^{(g-1)(1-\delta_{n,0})+1-\frac{d_{g,n}}{2}}PH_c(C^{SU(2)/\bbZ_2}_{g,n};q)\\
&=q^{(g-1)(1-\delta_{n,0})+1}\bbH_{g,n}(0,\sqrt{q}).
\end{split}
\end{equation} 
This is the polynomial used in conjecture \ref{conj:chavar}.

\emph{Remark: }the Cauchy function $\Omega(z,w)$ \eqref{eq:Cauchy} is very similar to the Macdonald index \cite{Gadde:2011uv} of $T_N$ theory with $N\rightarrow \infty$. We do not know the reason behind this similarity yet.

\section{Main results}
\label{sec:results}

Now we are ready to state our results. First consider cases with $n=0$, since there is no flavor symmetry from punctures, the highest weight state of a simple module $L_a$ is determined by its conformal dimension $h(L_a)$ \footnote{There is an $U(1)_f$ flavor symmetry discussed in \cite{Beem:2013sza} that is not visible in the class-$\mathcal{S}$ construction. Additionally, there is an additional accidental $USp(4)$ symmetry at the level of associated VOA \cite{Kiyoshige:2020uqz,Beem:2021jnm}. However, in this paper we will only consider flavor symmetry from the punctures in the class-$\mathcal{S}$ construction. It would be interesting to explore how the additional symmetry further refines the our proposal.}, and all $L_a$'s are ordinary modules. We have

\begin{conj}\label{conjn0}When $n=0$, there is a bijection between the $g$ simple modules $\{L_a\}$ of $V_{g,0}$ and the $g$ fixed manifolds $\{M_a\}_{0\leq a\leq g-1}$ of $\CM_{g,0}$ such that
\begin{equation}
\label{eq:bijectionn0}
h(L_a)=\mu(M_a) -g+\frac{3}{2}-\frac{1}{2}\delta_{\mu(M_a),0} = \mu - \mu_\text{max} - \frac{1}{2}\delta_{\mu(M_0), 0} \ .
\end{equation}
Moreover, the Jordan type of modular $T$ matrix is 
\begin{equation}
[(1-\delta_{\mu(M_a),0})(\dim M_a+1) +g]_{M_a\in \CM_{g,0}^T},
\end{equation}
with $g$ Jordan blocks.
\end{conj}
The order of MDE (\ref{eq:MDEofT}) of the theory is just the sum of sizes of all Jordan blocks of $T$. Our results compute the order of MDE in terms of dimensions of fixed manifolds.

When adding regular punctures on the Riemann surface, one needs also assign a parabolic parameter $\alpha_i$ for the $i$-th puncture. The value of moment map $\mu$ on each fixed manifold will be a function of the parameters $\{\alpha_i\}_{1\leq i\leq n}$. The correspondence in this case is given by the following.
\begin{conj}
\label{conj2}
When $n>0$ and assuming $0<\alpha_n<\cdots<\alpha_1<1/4$, there is a bijection between the simple modules $\{L_a\}$ of $V_{g,n}$ and the fixed manifolds $\CM^T_{g,n} = \{M_a\}$ such that, for even $n$,
\begin{eqnarray}
\label{eq:bijneven}\mu_{M_a} - \mu_\text{max}|_{\alpha_i\mapsto -\omega^{(i)}}
= h(L_a) +\sum_{i=1}^n \lambda_i \omega^{(i)},
\end{eqnarray} 
and for odd $n$,
\begin{equation}
  \label{eq:bijnodd} \mu_{M_a} - \mu_\text{max}|_{\alpha_1 \mapsto \frac{1}{2} + \omega^{(1)}, \alpha_{i > 1}\mapsto \frac{1}{2}-\omega^{(i)}}
  = h(L_a) +\sum_{i=1}^n \lambda_i \omega^{(i)}
\end{equation}
Here $\mu_{max}$ is defined in section \ref{sec:hitchin}, $\omega^{(i)}$ is the fundamental weight of the $i$-th flavor $\fsu(2)$. The highest weight state of $L_a$ has conformal dimension $h(L_a)$ and transforms in the representation $[\lambda_1,\lambda_2,\cdots,\lambda_n]$ under the flavor symmetry $\fsu(2)^{(1)}\times\cdots \fsu(2)^{(n)}$. Moreover, the Jordan type of modular $T$ (resp. $STS$) is 
\begin{equation}
\label{eq:dimJord}
[(1-\delta_{\mu(M_a),0})(\dim M_a+1) +g]_{M_a\in \CM_{g,n}^T},
\end{equation}
for $n$ even (resp. odd).
\end{conj}
When $g=0$ the total number of simple and logarithmic modules is the same as the dimension of the cohomology of $\CM_{0,n}$, hinting strongly that logarithmic modules correspond to elements in the cohomology ring of $\CM_{0,n}$.

As the moment map of an fixed manifold completely determines the highest weight state of the corresponding simple module, we can identify all the ordinary modules, and compute the modular matrices among them (and some of the logarithmic modules) such as $T^\text{ord}$, $(STS)^\text{ord}$.
\begin{cor}
  \label{corollary1}All ordinary modules of $V_{g,n}$ correspond to fixed manifolds in the subset
	\begin{equation}
		\CM^\text{ord}_{g,n} = \left\{\begin{array}{ll}
			\{M_{d,(1,1,\cdots,1)}\}_{-n+\sum_i\alpha_i<d\leq g-1-n/2} \ , & \text{if $n$ is even} ,\\
			\\
			\{M_{d,(0,1,\cdots,1)}\}_{-n+1-\alpha_1+\sum_{i>1}\alpha_i<d\leq g-1-(n-1)/2} \ , & \text{if $n$ is odd},
		\end{array}
		\right.
	\end{equation}
such that
\begin{equation}
	\mu(M)-\mu_{max} = h(L^\text{ord}).
\end{equation}
The Jordan type of $T^\text{ord}$ (resp. $(STS)^\text{ord}$) is $[\dim M+g+2]_{M\in \CM^\text{ord}_{g,n}}$ (resp.  $[\dim M+g+1]_{M \in \CM^\text{ord}_{g,n}}$).
\end{cor}
Again, the total size of $T^\text{ord}$ or $(STS)^\text{ord}$ (hence the order of MDE) can be computed using the dimensions of fixed manifolds corresponding to ordinary modules.

As mentioned before $\CM_{g,n}$ is homeomorphic to the character variety $C_{g,n}$. We also find the following  relations among $C_{g,n}$, $\CM_{g,n}$  and $V_{g,n}$.
\begin{conj}\label{conj:chavar}Let $P_{g,n}(q)= \sum_{i}a_i q^{d_i} $ be the renormalization of the pure part of the mixed Hodge polynomial of $\CM^{SU(2)/\bbZ_2}_{g,n}$ defined in \eqref{eq:renormalizedMHP}. The total number of fixed manifolds of $\CM_{g,n}$ (i.e. number of simple modules of $V_{g,n}$) is $P_{g,n}(1)=\sum_i a_i$, $\dim \bbV^{mod}_{g,n} = \sum_i a_i d_i$, and the Jordan type of the modular matrix is $[d_i^{a_i}]$. 
\end{conj}
This conjecture, similar to $P=W$ conjecture, also builds a bridge between renormalized mixed Hodge polynomials of character varieties and fixed manifolds of Hitchin moduli spaces. To the best of our knowledge, this correspondence was not noticed in math literature. 

\section{Examples}
\label{sec:examples}

\subsection{\texorpdfstring{$g\geq2, n=0$ cases}{}}

As there is no flavor symmetry from the punctures when $n=0$, in this subsection only ordinary modules and the corresponding logarithmic sector will be considered. As a result, we will focus on the MDEs in the $n = 0$ cases.

{\bf Example $(g,n)=(2,0)$:} The associated VOA has central charge $c = -26$, which has been studied extensively \cite{Kiyoshige:2020uqz,Beem:2021jnm}.
The vacuum character $\ch_0$ satisfies a sixth order MDE given in \cite{Beem:2017ooy},
\begin{align}
 \Big[D_q^{(6)} - & \ 305 E_4 D_q^{(4)} - 4060E_6 D_q^{(3)}
      + 20275E_4^2 D_q^{(2)} \nonumber \\
      & \ + 2100E_4 E_6 D_q^{(1)}  - (68600E_6^2 - 49125E_4^3) \Big]\ch_0 =0 .
\end{align}
with two non-log solutions
\begin{equation}
  \ch_0 = \frac{1}{2}\eta(\tau)^2(E_2 + \frac{1}{12}), \qquad
  \ch_1 = \eta(\tau)^2 \ ,
\end{equation}
which correspond to the vacuum module ($h=0$) and another module with $h=-1$ respectively.
There are also four log solutions,
\begin{align}
  \ch^{\text{long}}_1 = & \ - \frac{1}{24} i \tau \eta(\tau)^2 + \frac{1}{4\pi} \tau^2 \eta(\tau)^2 - \frac{i}{2}\tau^3 \eta(\tau)^2 E_2(\tau) \\
  \ch^{\text{long}}_2 = & \ - \frac{1}{24} i \tau \eta(\tau)^2 + \frac{1}{4\pi} \eta(\tau)^2 + \frac{1}{2\pi} \tau \eta(\tau)^2 - \frac{i}{2} \eta(\tau)^2 E_2(\tau)\\
  & \  - \frac{3}{2}i \tau \eta(\tau)^2 E_2(\tau) - \frac{3}{2}i \tau^2 \eta(\tau)^2 E_2(\tau)\\
  \ch^{\text{long}}_3 = & \ + \frac{1}{2\pi}\eta(\tau)^2 - 3i \eta(\tau)^2 E_2 - 3i\tau\eta(\tau)^2 E_2(\tau)\\
  \ch^{\text{long}}_4 = & \ \frac{i}{\pi}\tau \eta(\tau)^2 - \frac{3i}{2\pi} \tau^2 \eta(\tau)^2 + 3 \tau^2 \eta(\tau)^2 E_2(\tau) - 3 \tau^3 \eta(\tau)^2 E_2(\tau) \ .
\end{align}
where as usual $q = e^{2\pi i \tau}$. These six solutions span a six-dimensional representation of the modular group $SL(2,\bbZ)$. In the ordered basis $(\ch_0, \ch_1, \ch^\text{log}_3, \ch^\text{log}_2, \ch^\text{log}_4, \ch^\text{log}_1)$, the matrix representation of the $T$ transformation is
\begin{equation}
  \left(
    \begin{array}{cccccc}
     e^{\pi i/6} & 0 & 0 & 0 & 0 & 0 \\
     0 & e^{\pi i/6} & 0 & 0 & 0 & 0 \\
     -6 e^{2\pi i/3} & \frac{1}{2} e^{2\pi i/3} & e^{\pi i/6} & 0 & 0 & 0 \\
     0 & 0 & e^{\pi i/6} & e^{\pi i/6} & 0 & 0 \\
     -6 e^{\pi i/6} & e^{\pi i/6} & 2 e^{2\pi i/3} & -6 e^{2\pi i/3} & e^{\pi i/6} & 0 \\
     0 & 0 & 0 & e^{\pi i/6} & 0 & e^{\pi i/6} \\
    \end{array}
    \right)
\end{equation}
which has Jordan type $[2,4]$, the same as the prediction from the dimension of fixed manifolds and all diagonal elements in the Jordan normal form are $e^{\pi i/6}$. The renormalized mixed Hodge polynomial of $C^{SU(2)/\bbZ_2}_{2,0}$ is 
\begin{equation}
P_{2,0}(q)=q^2+q^4.
\end{equation} 
Therefore our proposal is confirmed in this example.

{\bf General $g \ge 2$, $n = 0$: } The associated VOA $V_{g, 0}$ has not been explicitly constructed. However, we may resort to indices of $\CT_{g,0}$ to make predictions on the representation theory of $V_{g, 0}$.
The vortex indices $\CI^{\text{vortex}}_{g,0}(\kappa)$ with even vorticity $\kappa$ are linear combinations of $\eta(\tau)^{2g - 2}E_{2\ell}(\tau), \ell = 0, 1, 2, ..., g - 1$, which predicts $g$ ordinary modules $L(h)$ with conformal weights\footnote{For Argyres-Douglas theories, these vortex type defects do correspond to irreducible modules of the associated VOA. However, we are unable to show the modules here to be irreducible. Nevertheless, we proceed by assuming irreducibility.},
\begin{equation}
h = 0, -1, -2, ... \ , - g + 1 \ .
\end{equation}
Here $h = 0$ corresponds to $\kappa = 0$, the vacuum module of the VOA.

Let us denote $h_d = d - g + 1$ with $d \coloneqq (g - 1) - \frac{\kappa}{2}$ given in terms of the vorticity $\kappa$. We have a bijection between modules and fixed manifolds
\begin{equation}
	L(h_d) \leftrightarrow M_d \ .
\end{equation}
The relation between the conformal weights and critical values of the moment map is simply
\begin{equation}
h_d	= \mu(M_d) - \mu_\text{max} - \frac{1}{2}\delta_{d, 0} \ ,
\end{equation}
where $\mu(M_{d > 0}) = d - \frac{1}{2}$, $\mu(M_0) = 0$, and $\mu_\text{max} = g - \frac{3}{2}$. In particular, we see that the vacuum module $L(0)$ corresponds to the fixed manifold $M_{g - 1}$, while the module $L(h = -g + 1)$ associated to the vorticity $\kappa = 2g - 2$ corresponds to the fixed manifold $M_0$. The Kronecker delta term compensates the jump in moment map $\mu(M_d)$ from $d = 0$ to $d > 0$. This relation is consistent with the relation in \cite{Fredrickson:2017yka,Fredrickson:2017jcf} as there are no $M_0$ when $g=0$.

%

Using either  vortex indices or $\{\eta(\tau)^{2g - 2}E_{2\ell}\}$, we compute the modular $T$ matrix. The Jordan type of the modular $T$ matrix is $[g,g+2,g+4,\cdots, 3g-2]$, where the sizes of the blocks can be identified with $[g, \dim M_{g - 1} + g + 1, ..., \dim M_1 + g + 1]$. The total dimension of $\bbV$ (i.e. the total order of  MLDE) is then
\begin{equation}
g+\sum_{d=1}^{g-1}(\dim M_d +g +1)=g(2g-1),
\end{equation}
matching the conjecture in \cite{Beem:2021zvt}. Using equation \ref{eq:mhpg0} one finds the renormalized mixed Hodge polynomial of $C^{SU(2)/\bbZ_2}_{g,0}$ is
\begin{equation}
P_{g,0}(q)=q^g+q^{g+2}+q^{g+4}+\cdots+q^{3g-2},
\end{equation}
which confirms conjecture \ref{conj:chavar} in the case when $n=0$.

\subsection{$g, n > 0$ cases}

When the number $n$ of punctures is non-zero, the four dimensional $A_1$ theory enjoys at least an $SU(2)^n$ flavor symmetry, implying the existence of $n$ affine $\widehat{\mathfrak{su}}(2)_{-2}$ subalgebras in $V_{g, n}$. The flavor symmetry acts non-trivially on the Higgs branch which corresponds to the associated variety $X_{V_{g,n}}$. Since the classifications of simple modules for most $V_{g,n}$ are unknown, we will use method described in section \ref{sec:modularmodules} to study its ordinary, simple and log modules, and match with fixed manifold data. Our results also provide predictions on classifications of simple modules of $V_{g,n}$.



\vspace{1em}

{\bf Example $g=0$, $n=3$}

The theory $\mathcal{T}_{0,3}$ is the building block of the $A_1$ class-$\mathcal{S}$ theories. It is the theory of four free hypermultiplets which carries a $USp(8)$ flavor symmetry, where only the $SU(2)^3$ part is visible in the class-$\mathcal{S}$ construction. The associated VOA $V_{0,3}$ is four decoupled $\beta \gamma$ systems. The Schur index is 
\begin{equation}
	\mathcal{I}_{0,3} = \prod_{\alpha, \beta = \pm} \frac{\eta(\tau)}{\vartheta_1(\mathfrak{b}_1 + \alpha \mathfrak{b}_2 + \beta \mathfrak{b}_3)} \ .
\end{equation}
There are two crucial flavored MLDEs coming from the null states of the chiral algebra,
\begin{align}
  \left[D_q^{(1)} - \frac{1}{2} \sum_{\alpha_i = \pm}E_2 \begin{bmatrix}
      	-1 \\ \prod_{i = 1}^{3}b_i^{\alpha_i}  
    	\end{bmatrix}\right]\mathcal{I}_{0,3} = & \ 0\\
  \left[D_{b_i} - \sum_{\alpha, \beta = \pm}E_1 \begin{bmatrix}
      	-1 \\ b_i b_j^\alpha b_k^\beta
    	\end{bmatrix}\right]\mathcal{I}_{0,3} = & \ 0, \quad i \ne j \ne k
\end{align}
It straightforward to show that the two equations have only one simultaneous solution given simply by $\mathcal{I}_{0,3}$.

On the Hitchin moduli space side, there is only one fixed manifold $M_{-2, (0,1,1)}$ with the unique moment map $\mu - \mu_\text{max} = 0$. The fixed manifold corresponds precisely to the unique character $\mathcal{I}_{0,3}$.

\vspace{1em}
{\bf Example $g=0$, $n=4$}

$\CT_{0,4}$ is the $\mathcal{N} = 2$ $SU(2)$ gauge theory coupled with four fundamental hypermultiplets. The associated VOA $V_{0,4}$ is isomorphic to the affine vertex algebra $L_{-2}(D_4)$ whose simple modules are classified in \cite{pervse2013note, Arakawa_2016}. There are four non-vacuum simple modules whose highest weight states have $D_4$ weights $-2 \omega_1, - \omega_2, -2 \omega_3, -2 \omega_4$. If we choose $\{\theta, \alpha_1, \alpha_3,\alpha_4\}$ of $D_4$ to be  positive roots of the $\fsu(2)^{\otimes 4}$ subalgebra of $D_4$, Dynkin labels of highest weight states of non-vacuum simple modules under $\fsu(2)^{\otimes 4}$ will be $[-2,-2,0,0]$, $[-2,0,0,0]$, $[-2,0,-2,0]$, and $[-2,0,0,-2]$. The vacuum module has $0$ conformal weight while other four simple modules have conformal weight $-1$.

\begin{table}[h]
  \normalsize
  \begin{tabular}{c|c||c|c|c|c}
  $L(\Lambda)$ & $h(\Lambda)$ & $ F(\Lambda) = M_{d,e}$ & $\dim$  & $\mu(F(\Lambda))$ & $\mu-\mu_\text{max}$ \\ \hline
  $L(-\omega_2)$ & $-1$ & $M_{-3,\{0,1,1,1\}}$ & $1$ & $\alpha_1-\alpha_2-\alpha_3-\alpha_4$ & $-1+2\alpha_1$ \\ \hline
  $L(-2\omega_1)$ & $-1$  &$M_{-2,\{0,0,1,1\}}$ & $0$ & $\alpha_1+\alpha_2-\alpha_3-\alpha_4$ & $-1+2\alpha_1+2\alpha_2$  \\
  $L(-2\omega_3)$ & $-1$ & $M_{-2,\{0,1,0,1\}}$ & $0$ & $\alpha_1-\alpha_2+\alpha_3-\alpha_4$ &  $-1+2\alpha_1+2\alpha_3$  \\
  $L(-2\omega_4)$ & $-1$ & $M_{-2,\{0,1,1,0\}}$ & $0$ & $\alpha_1-\alpha_2-\alpha_3+\alpha_4$ &  $-1+2\alpha_1+2\alpha_4$ \\
  $L(-2\omega_0)$ & $0$ & $M_{-3,\{1,1,1,1\}}$ & $0$ & $1-\alpha_1-\alpha_2-\alpha_3-\alpha_4$ & $0$
  \end{tabular}
  \caption{\label{table:g0n4fps}Fixed points for $g=0$, $n=4$. Here $F(\Lambda)$ represents the fixed manifold corresponds to the simple module $L(\Lambda)$.}
\end{table}

There are five fixed manifolds in the Hitchin moduli space. The data of the five fixed manifolds are collected in Table \ref{table:g0n4fps}, and the bijection between simple modules and  fixed manifolds is clear from the same table. The maximal value $\mu_\text{max} = 1 - \sum_{i=1}^{4} \alpha_i$ of moment map is defined in section \ref{sec:hitchin}, namely, the moment map on the fixed manifold $M_{-3, \{1,1,1,1\}}$ corresponding to the vacuum module. $\mu-\mu_{max}$ of other four fixed manifolds are $-1+2\alpha_1$, $-1+2\alpha_1+2\alpha_2$, $-1+2\alpha_1+2\alpha_3$, and $-1+2\alpha_1+2\alpha_4$. Under the change of variable $\alpha_i\mapsto -\omega^{(i)}$ they become
\begin{equation}
\mu-\mu_{max}|_{\alpha_i\mapsto \omega^{(i)}}:\ -1-2\omega^{(1)},\ -1-2\omega^{(1)}-2\omega^{(2)},\ -1-2\omega^{(1)}-2\omega^{(3)},\ -1-2\omega^{(1)}-2\omega^{(4)}.
\end{equation}
Clearly the constant term gives $h$ and coefficients of $\omega^{(i)}$'s give the Dynkin label of the corresponding module.

Characters of all simple modules are linear combinations of residues $\operatorname{R}_i \coloneqq \operatorname{Res}_{a \to m_i} \mathcal{Z}(a)$ \cite{Zheng:2022zkm,2023arXiv230409681L},
\begin{align}
  \ch_{- 2\omega_1} = & \ \mathcal{I}_{0,4} - 2\operatorname{R}_1\\
  \ch_{- \omega_2} = & \ - \mathcal{I}_{0,4} + 2\operatorname{R}_1 + 2\operatorname{R}_2 \\
  \ch_{- 2\omega_3} = & \ \mathcal{I}_{0,4} - \operatorname{R}_1 - \operatorname{R}_2 - \operatorname{R}_3 - \operatorname{R}_4 \\
  \ch_{- 2\omega_4} = & \ \mathcal{I}_{0,4} - \operatorname{R}_1 - \operatorname{R}_2 - \operatorname{R}_3 + \operatorname{R}_4 \ ,
\end{align}
where $\mathcal{Z}(a)$ denotes the integrand that computes the Schur index $\mathcal{I}_{0,4}$ via a contour integral $\mathcal{I}_{0,4} = \oint \mathcal{Z}(a)$. The characters of the vacuum, above simple modules and an extra logarithmic solution have been shown to be the complete solution set to the FMDEs coming from the null states in $L_{-2}(D_4)$ \cite{Zheng:2022zkm,Pan:2023jjw}, so the dimension of $\bbV^{\mathrm{mod}}_{0,4}$ is $6$ and the Jordan type of $T$ matrix is $[2,1^4]$.
The  Jordan type of $T$  matches exactly  $\dim M+1$ from each fixed manifold. The renormalized mixed Hodge polynomial is
\begin{equation}
P_{0,4}(q)= 4q + q^2,
\end{equation}
reflecting the Jordan type of $T$, as consistent with the conjecture.

\vspace{1em}
{\bf Example $g=1$, $n=1$}

$\CT_{1,1}$ is the $\mathcal{N}=4$ $SU(2)$ super Yang-Mills coupled with a free hypermultiplet. The associated VOA $V_{1,1}$ is  isomorphic to the $2d$ small $\CN=4$ superconformal algebra with central charge $c = -9$ coupled with a free $\beta \gamma$ system with $c = -1$ that descends from the free hypermultplet \cite{Beem:2013sza}. For simplicity, we only turn on the diagonal $SU(2)$ of the $SU(2) \times SU(2)$ flavor symmetry. As classified in \cite{Adamovic:2014lra}, there are only two simple modules from category-$\mathcal{O}$ of the 2d small $\mathcal{N} = 4$ superconformal algebra. On the other hand, there are three independent FMDEs, of weight-two, three and four respectively, with  two non-log solutions of weight $h = 0$ and $h = - \frac{1}{2}$,
\begin{equation}
  \frac{\eta(\tau)}{\vartheta_1(2\mathfrak{b}_1)}E_1 \begin{bmatrix}
    -1 \\ b_1
  \end{bmatrix}, \qquad \frac{\eta(\tau)}{\vartheta_1(2\mathfrak{b}_1)}\ .
\end{equation}
The first solution, which is simply the Schur index without defect, belong to a two dimensional representation of the modular group $\Gamma^0(2)$. The second solution corresponds to the character of a free field realization of $V_{1,1}$ in terms of a $bc\beta \gamma$ system \cite{Bonetti:2018fqz} coupled with the $c = -1$ free $\beta \gamma$ system, which corresponds to the module $L(-\omega_1)$ in \cite{Adamovic:2014lra} (ignoring the factor of vacuum module of the $\beta \gamma$ system). $L(-\omega_1)$ is non-ordinary, whose $h = -1/2$ weight space forms an infinite dimensional representation of $\fsu(2)$ with the highest weight $-\omega_1$. This result implies that of $\bbV_{1,1}$ is $3$ dimensional. The $T^2$ matrix can be easily computed with  Jordan type $[2,1]$.

On the Hitchin  side, WLOG, there are two fixed manifolds, $M_0$ and $M_{0,0}$. The moment map $\mu$ are summarized in Table \ref{table:g1n1fps-2-1}. 
\begin{table}[h]
\begin{tabular}{c|c||c|c|c}
$h$ &$\fsu(2)$ rep &  $M_{d,e}$ & $\dim$  &   $(\mu-\mu_{max})|_{\alpha\mapsto \frac{1}{2}+\omega}$ \\ \hline
$-1/2$ & $[-1]$   &$M_{0}$ & $-$  &$-1/2-\omega$  \\ \hline
$0$ & $[0]$ & $M_{0,0}$ & $0$  & $0$ 
\end{tabular}
\caption{\label{table:g1n1fps-2-1}$g=1$, $n=1$.  $\dim M_{0,0}+g+1=2$ so it corresponds to the $2\times 2$ Jordan block of $T^2$. We are not listing the dimension of $M_0$ because it is not used in all conjectures.}
\end{table}
%
%
The bijection between  simple modules and  two fixed manifolds is obvious, and again the constant term and the coefficient of $\omega$ in $\mu-\mu_{max}|_{\alpha\mapsto \frac{1}{2}+\omega}$ gives conformal weight and Dynkin labels respectively. The renormalized mixed Hodge polynomial is
\begin{equation}
P_{1,1}(q) = q+q^2,
\end{equation}
which, together with dimensions of fixed manifolds,  encodes the Jordan type of $T^2$.

\vspace{1em}
{\bf Example $(g,n)=(1,2)$}

$\CT_{1,2}$ is given by an $SU(2)^2$ gauge theory coupled with some fundamental and adjoint hypermultiplets. Little is known about the associated VOA $V_{1,2}$, therefore fixed manifold data of Hitchin side predicts  the the representation theory of $V_{1,2}$. The four fixed manifolds in $\CM_{1,2}$ are summarized in table \ref{table:g1n2fps}, and the renormalized mixed Hodge polynomial is
\begin{equation}
P_{1,2}(q)=q(1+2q+q^2),
\end{equation}
so we predict that there are $4$ simple modules of $V_{1,2}$, the dimension of $\bbV_{1,2}$ is $8$, and the Jordan type of modular $T$-matrix is $[3,2^2,1]$.
\begin{table}[h]
  \begin{tabular}{c|c||c|c|c}
		$h$ & $\fsu(2)^{\otimes 2}$ Dynkin label & $M_{d,e}$ & $\dim$  & $\mu-\mu_{max}|_{\alpha_i\mapsto -\omega^{(i)} }$\\ \hline
		$-1$ & $[-1,-1]$ & $M_{0}$ & $-$ & $-1-\omega^{(1)}-\omega^{(2)}$\\\hline
		$-1$ & $[-2,-2]$ & $M_{0,(0,0)}$ & $0$ & $-1-2\omega^{(1)}-2\omega^{(2)}$\\ \hline
		$-1$ & $[-2,0]$ & $M_{-1,(0,1)}$ & $1$ & $-1-2\omega^{(1)}$\\ \hline
		$0$ &$[0,0]$ & $M_{-1,(1,1)}$ & $0$ & $0$
  \end{tabular}
  \caption{\label{table:g1n2fps}$g=1$, $n=2$. The conformal dimension and Dynkin labels are the same as constant terms and coefficients of $\omega^{(i)}$ of $\mu-\mu_{max}|_{\alpha_i\mapsto -\omega^{(i)}}$. The dimension of $M_0$ is not listed because it is not used in all conjectures.}
\end{table}


As a consistency check, we compare predictions from Hitchin side with characters from indices. Since the work of \cite{Zheng:2022zkm}, several FMDEs up to modular-weight-five have been constructed. As reviewed in section \ref{sec:modularmodules}, solutions of these FMDEs can be constructed from computing the Schur index with or without defects \cite{Guo:2023mkn}.
\begin{figure}
  \centering
  \includegraphics{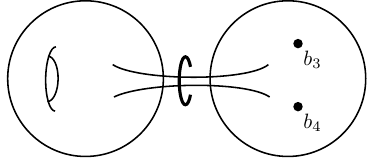}
  \caption{A Wilson line operator in the genus-one theory with two punctures. Each long tube or handle denotes an $SU(2)$ gauge group.} \label{fig:genus-one-type-2}
\end{figure}
For example, one may insert a half Wilson line operator transforming in the spin-$j$ representation with respect to the $SU(2)$ gauge group shown in figure \ref{fig:genus-one-type-2}, and the line index $\langle W_j\rangle$
\begin{align}
  \langle W_j\rangle = & \ \delta_{j \in \mathbb{Z}} \mathcal{I}_{1,2}
  - \frac{\eta(\tau)^2}{4\prod_{i = 1}^{2}\vartheta_1(2 \mathfrak{b}_i)}
  \sum_{\substack{m = -j\\ m \ne 0}}^{+j} (q^m + q^{-m}) \prod_{i = 1}^{2}\frac{b_i^{2m} - b_i^{-2m}}{q^m - q^{-m}} \nonumber\\
  & \ - \frac{\eta(\tau)^2}{2 \prod_{i = 1}^{2}\vartheta_1(2 \mathfrak{b}_i)}\sum_{\alpha = \pm}\Big(
    \alpha E_1 \begin{bmatrix}
      1 \\ b_1 b_2^\alpha
    \end{bmatrix}
    \sum_{\substack{m = -j\\ m\ne 0}}^{+j}\frac{(b_1b_2^\alpha)^{2m} - (b_1b_2^\alpha)^{-2m}}{q^m - q^{-m}}
    \Big)  \ .
\end{align}
is the linear combination of the following four quasi-modular forms including the Schur index $\CI_{1,2}$
\begin{equation}
\label{eq:basisV12}
  \mathcal{I}_{1,2}, \quad \frac{\eta(\tau)^{2}}{\prod_{i = 1}^{2}\vartheta_1(2 \mathfrak{b}_i)}, \quad 
  \frac{\eta(\tau)^{2}}{\prod_{i = 1}^{2}\vartheta_1(2 \mathfrak{b}_i)} E_1 \begin{bmatrix}
    1 \\ b_1 b_2
  \end{bmatrix}, \quad 
  \frac{\eta(\tau)^{2}}{\prod_{i = 1}^{2}\vartheta_1(2 \mathfrak{b}_i)} E_1 \begin{bmatrix}
    1 \\ b_1 b_2^{-1}
  \end{bmatrix} \,
\end{equation}
with coefficients being rational functions of $b_i$ and $q$. These four expressions are solutions to all the known FMDEs, and are convincing candidates of module characters. 
They be recombined into four $q$-series which we believe be the characters of simple modules of $V_{1,2}$.
\begin{align}
	\mathcal{I}_{1,2} = & \ q^{5/6} + \cdots \ ,  \\
	\frac{\eta(\tau)^2}{\prod_{i = 1}^{2}\vartheta_1(2 \mathfrak{b}_i)} = & \ \frac{b_1^{-1} b_2^{-1}}{(1 - b_1^{-2}) (1 - b_2^{-2})} q^{-1/6} + \cdots ,  \\
	 \frac{\eta(\tau)^2}{\prod_{i = 1}^{2}\vartheta_1(2 \mathfrak{b}_i)} \left(E_1 \begin{bmatrix}
	 	   	1 \\ b_1^{-1}b_2^{-1}
	 	 	\end{bmatrix} + \frac{1}{2}\right) = & \ \frac{b_1^{-2} b_2^{-2} q^{-1/6}}{(1 - b_1^{-2})(1 - b_2^{-2})(1 - b_1^{-1} b_2^{-1})} + \cdots \ , \\
	 \frac{\eta(\tau)^2}{\prod_{i = 1}^{2}\vartheta_1(2 \mathfrak{b}_i)} \left(E_1 \begin{bmatrix}
	 	   	1 \\ b_1^{-1}b_2  
	 	 	\end{bmatrix} + \frac{1}{2}\right) = & \ \frac{b_1^{-2} q^{-1/6} }{(1 - b_1^{-2})(1 - b_2^{-2})(1 - b_1^{-1} b_2)}+ \cdots \ . 
\end{align}
Here we have used $E_k\big[\substack{+1\\b^{-1}}\big] = (-1)^k E_k \big[\substack{+1\\b}\big]$. From these $q$-series expansions, we can read off the conformal weight $h = 0, -1, -1,-1$, and the $\fsu(2)^2$ weights $0$, $-\omega^{(1)} - \omega^{(2)}$, $-2 \omega^{(1)} - 2 \omega^{(2)}$, $-2\omega^{(1)}$ of corresponding modules. Here $\omega^{(i)}$ denotes the fundamental weight of the $i$-th $\fsu(2)$. These data matches exactly the prediction from fixed manifolds.

Finally, the Schur index $\mathcal{I}_{1,2}$ and other quasi-modular forms in equation \ref{eq:basisV12} belong to a dimension $8$ representation of the modular group $SL(2, \mathbb{Z})$, and the $T$ matrix has the Jordan type $[3, 2^2, 1]$, again consistent with the prediction from Hitchin side and renormalized mixed Hodge polynomials.

\vspace{1em}
{\bf Example $(g,n)=(g \ge 2,1)$}

Little is known about the associated VOA of these theories, so we again extract module information from indices and compare it with fixed manifolds data from Hitchin side.
One can show that the Schur index $\CI_{g,1}$, the vortex surface defect indices $\CT^{\mathrm{vortex}}_{g,1}(\kappa=\mathrm{even})$, and Wilson line indices are linear combinations of 
\begin{equation}
	\frac{\eta(\tau)^{2g - 2 + n}}{\vartheta_1(2 \mathfrak{b}_1)}, \quad 
	\frac{\eta(\tau)^{2g - 2 + n}}{\vartheta_1(2 \mathfrak{b}_1)} E_k \begin{bmatrix}
  	-1 \\ b_1  
	\end{bmatrix}, \quad k = 1, 2, \cdots, 2g - 2 + 1 \ ,
\end{equation}
therefore they are likely to form a basis of characters of simple modules.

As a simplest example, we consider $g = 2$. This basis of indices consists of four quasi-modular forms,
\begin{equation}
	\frac{\eta(\tau)^3}{\vartheta_1(2 \mathfrak{b}_1)}, \quad
	\frac{\eta(\tau)^3}{\vartheta_1(2 \mathfrak{b}_1)} E_1 \begin{bmatrix}
    -1 \\ b_1
	\end{bmatrix}, \quad
	\frac{\eta(\tau)^3}{\vartheta_1(2 \mathfrak{b}_1)} E_2 \begin{bmatrix}
    -1 \\ b_1
	\end{bmatrix}, \quad
	\frac{\eta(\tau)^3}{\vartheta_1(2 \mathfrak{b}_1)} E_3 \begin{bmatrix}
    -1 \\ b_1
	\end{bmatrix}\ ,
\end{equation}
which can be used to construct the following four $q-$series which should be characters of simple modules,
\begin{align}
	\frac{i \eta(\tau)^3}{\vartheta_1(2 \mathfrak{b}_1)} = & \ \frac{b_1^{-1}}{1 - b_1^{-2}} + O(q) \\
  \mathcal{I}^\text{vortex}_{2,1}(2) = \frac{i\eta(\tau)^3}{\vartheta_1(2 \mathfrak{b}_1)} \left(
  \frac{5}{8}E_1 \begin{bmatrix}
  	-1 \\ b_1  
	\end{bmatrix} + 3 E_3 \begin{bmatrix}
  	-1 \\ b_1  
	\end{bmatrix}
  \right) = & \ q^{\frac{1}{2}} + O(q^{\frac{3}{2}})\\ 
	\frac{i \eta(\tau)^3}{\vartheta_1(2 \mathfrak{b}_1)} \left(E_2 \begin{bmatrix}
  	-1 \\ b_1  
	\end{bmatrix} + \frac{1}{2}E_1 \begin{bmatrix}
  	-1 \\ b_1  
	\end{bmatrix} - \frac{1}{24}\right) = & \ \frac{b_1^{-2}q^{\frac{1}{2}}}{1 - b_1^{-2}} + O(q^{\frac{3}{2}}) \\ 
	\mathcal{I}_{2,1} = \frac{i\eta(\tau)^3}{\vartheta_1(2 \mathfrak{b}_1)}\left(
	- \frac{1}{8}E_1 \begin{bmatrix}
  	-1 \\ b_1  
	\end{bmatrix}
	+ E_3 \begin{bmatrix}
  	-1 \\ b_1  
	\end{bmatrix}
	\right) = & \ q^{\frac{3}{2}} + O(q^{\frac{5}{2}}) \\ 
\end{align}
From the leading term of each character, we predicts  conformal weights $h$ and Dynkin labels of four simple modules which are summarized in table \ref{table:V21}.
We further perform modular transformation on the basis to obtain the full basis of $\bbV_{2,1}$ and explicit expressions of  $STS, T^2$ matrices. In the end, $\bbV_{2,1}$ is $14$ dimensional and  the Jordan type of $T^2$ is $[2, 3, 4, 5]$. They are the same as predictions from fixed manifolds which are also listed in table \ref{table:V21}. The renormalized mixed Hodge polynomial is
\begin{equation}
P_{2,1}(q)=q^2(1+q+q^2+q^3).
\end{equation}
\begin{table}[h]
  \begin{tabular}{c|c||c|c|c}
		$h$ & $\fsu(2)$ Dynkin label & $M_{d,e}$ & $\dim$  & $\mu-\mu_{max}|_{\alpha \mapsto \frac{1}{2}+\omega }$\\ \hline
		$-\frac{3}{2}$ & $[-1]$ & $M_{0}$ & $-$ & $-\frac{3}{2}-\omega$\\\hline
		$-1$ & $[0]$ & $M_{0,(0)}$ & $2$ & $-1$\\ \hline
		$-1$ & $[-2]$ & $M_{0,(1)}$ & $1$ & $-1-2\omega$\\ \hline
		$0$ &$[0]$ & $M_{1,(0)}$ & $0$ & $0$
  \end{tabular}
\caption{\label{table:V21} $g=2,n=1$. The conformal dimension and Dynkin labels are the same as constant terms and coefficients of $\omega^{(i)}$ of $\mu-\mu_{max}|_{\alpha \mapsto \frac{1}{2}+\omega}$. The dimension of $M_0$ is not listed because it is not used in all conjectures.}
\end{table}

Next consider $g = 3$. The basis of indices has six quasi-modular forms
\begin{equation}
	\frac{\eta(\tau)^5}{\vartheta_1(2\mathfrak{b}_1)},\qquad
	\frac{\eta(\tau)^5}{\vartheta_1(2 \mathfrak{b}_1)}E_k \begin{bmatrix}
  	-1 \\ b_1  
	\end{bmatrix}, \qquad k = 1, 2, \cdots, 5 \,
\end{equation}
which can be recombined into the following $q$-series which could be characters of simple modules
\begin{align}
  & \frac{i\eta(\tau)^5}{\vartheta_1(2 \mathfrak{b}_1)} = \frac{b_1^{-1}q^{\frac{1}{12}}}{1 - b_1^{-2}} + O(q^{\frac{13}{12}}) \\
  & \mathcal{I}_{3,1}^\text{vortex}(4) = \frac{i\eta(\tau)^5}{\vartheta_1(2 \mathfrak{b}_1)} \left(5 E_5 \begin{bmatrix}
    -1 \\ b_1  
	\end{bmatrix} + \frac{95}{24} E_3 \begin{bmatrix}
  	-1 \\ b_1  
	\end{bmatrix}
	+ \frac{63}{128}E_1 \begin{bmatrix}
  	-1 \\ b_1  
	\end{bmatrix}\right) = q^{\frac{7}{12}} + O(q^{\frac{13}{12}}) \\
	& \frac{i \eta(\tau)^5}{\vartheta_1(2 \mathfrak{b}_1)} \left(
	E_2 \begin{bmatrix}
  	-1 \\ b_1  
	\end{bmatrix}
	+ \frac{1}{2} E_1 \begin{bmatrix}
  	-1 \\ b_1  
	\end{bmatrix}
	- \frac{1}{24}
	\right) = \frac{b_1^{-2}q^{\frac{7}{12}}}{1 - b_1^{-2}} + O(q^{\frac{13}{12}}) \\
	& \mathcal{I}^\text{vortex}_{3,1}(2) = \frac{i\eta(\tau)^5}{\vartheta_1(2\mathfrak{b}_1)} \left(
	3 E_5 \begin{bmatrix}
  	-1 \\ b_1  
	\end{bmatrix}
	+ \frac{3}{8}E_3 \begin{bmatrix}
  	-1 \\ b_1  
	\end{bmatrix}
	- \frac{7}{128}E_1 \begin{bmatrix}
  	-1 \\ b_1  
	\end{bmatrix}\right) = q^{\frac{19}{12}} + O(q^{\frac{31}{12}})  \\
	& \frac{i\eta(\tau)^5}{\vartheta_1(2 \mathfrak{b}_1)} \left(
	E_4\begin{bmatrix}
  	-1 \\ b_1  
	\end{bmatrix}
	+ \frac{1}{2}E_3 \begin{bmatrix}
  	-1 \\ b_1  
	\end{bmatrix}
	- \frac{1}{24}E_2 \begin{bmatrix}
  	-1 \\ b_1  
	\end{bmatrix}
	- \frac{1}{16}E_1 \begin{bmatrix}
  	-1 \\ b_1  
	\end{bmatrix}
	+ \frac{17}{5760}
	\right) =  \frac{b_1^{-2} q^{\frac{19}{12}}}{1 - b_1^{-2}} + O(q^{\frac{31}{12}}) \\
	& \mathcal{I}_{3,1} = \frac{i\eta(\tau)^5}{\vartheta_1(2 \mathfrak{b}_1)}\left(
	E_5 \begin{bmatrix}
  	-1 \\ b_1  
	\end{bmatrix}
	- \frac{5}{24} E_3 \begin{bmatrix}
  	-1 \\ b_1  
	\end{bmatrix}
	+ \frac{3}{128} E_1 \begin{bmatrix}
  	-1 \\ b_1  
	\end{bmatrix}
	\right) =  q^{\frac{31}{12}} + O(q^{\frac{43}{12}})
\end{align}
Again properties of highest weight vectors can be read off from leading order terms and are summarized in table \ref{table:V31}.
These characters belong to a $33$-dimensional representation of the modular group $\Gamma^0(2)$, and the Jordan type of $T^2$ is $[8, 7, 6, 5, 4, 3]$. The fixed manifolds information is also listed in table \ref{table:V31}, and the renormalized mixed Hodge polynomial is
\begin{equation}
P_{3,1}(q)=q^3(1+q+q^2+q^3+q^4+q^5).
\end{equation}
All conjectures are correct in this case.

\begin{table}[h]
\begin{tabular}{c|c||c|c|c}
		$h$ & $\fsu(2)$ Dynkin label & $M_{d,e}$ & $\dim$  & $\mu-\mu_{max}|_{\alpha \mapsto \frac{1}{2}+\omega }$\\ \hline
		$-\frac{5}{2}$ & $[-1]$ & $M_{0}$ & $-$ & $-\frac{5}{2}-\omega$\\\hline
		$-2$ & $[0]$ & $M_{0,(0)}$ & $4$ & $-2$\\ \hline
		$-2$ & $[-2]$ & $M_{0,(1)}$ & $3$ & $-2-2\omega$\\ \hline
		$-1$ & $[0]$ & $M_{1,(0)}$ & $2$ & $-1$\\ \hline
		$-1$ & $[-2]$ & $M_{1,(1)}$ & $1$ & $-1-2\omega$\\ \hline
		$0$ &$[0]$ & $M_{2,(0)}$ & $0$ & $0$
  \end{tabular}
%
	\caption{\label{table:V31}$g=3,n=1$. The conformal dimension and Dynkin labels are the same as constant terms and coefficients of $\omega^{(i)}$ of $\mu-\mu_{max}|_{\alpha \mapsto \frac{1}{2}+\omega}$. The dimension of $M_0$ is not listed because it is not used in all conjectures.}
\end{table}


For general $g \ge 2$, the fixed manifolds are listed in table \ref{table:Vg1}. The renormalized mixed Hodge polynomial $P_{g,1}(q)$ is
\begin{equation}
P_{g,1}(q)=q^g(1+q+\cdots+q^{2g-1}).
\end{equation}
\begin{table}[h]
	\begin{tabular}{c|c|c}
		$M$ & $\mu - \mu_\text{max} |_{\alpha \to \frac{1}{2} + \omega}$ & $\dim M$\\
		\hline
		$M_0$ & $- \frac{g - 1}{2} - \omega$ & $-$ \\
		\hline
		$M_{d, (0)},\ 0\leq d\leq g-1$ & $d+1 - g$ & $2g-2-2d$\\
		$M_{d, (1)},\ 0\leq d\leq g-2$ & $d+1 - g - 2 \omega$ & $2g - 3-2d$\\
	\end{tabular}
	\caption{\label{table:Vg1} Fixed manifolds of $\CM_{g,1}$. The dimension of $M_0$ is not listed because it is not used in all conjectures.}
\end{table}

It is natural to identify the original Schur index and the $g - 1$ vortex defect indices with  the subset $\{M_{d, (0)}\}_{0\leq d\leq g-1}$ of fixed manifolds. The module whose character is $i\eta(\tau)^{2g - 2}/\vartheta_1(2 \mathfrak{b}_1)$ is identified with $M_0$. 
Remaining fixed manifold  $\{M_{d, (1)}\}_{0\leq d\leq g-2}$ are matched with modules whose characters start with $E_{\text{even }k} \big[\substack{-1 \\ b_1}\big]$.

\subsection{Ordinary modules of $V_{g,n}$}

In this subsection we study the subset $M_{g,n}^{\text{ord}}$ of fixed manifolds of $\CM_{g,n}$ which corresponds to ordinary modules of $V_{g,n}$. Given an ordinary module $L$ of $V_{g,n}$, the weight space of $L$ at any conformal weight should be finite dimensional, hence the highest weight state of $L$ should generate a finite dimensional representation of $\fsu(2)^{\otimes n}$, i. e. its Dynkin label $[j_1,\cdots j_n]$ should be all non-negative. 

On the Hitchin side, when $n$ is even, the difference between the critical value $\mu_{d,e}$ of the fixed manifold $\CM_{d,e}$ and $\mu_{\max}$ is
\begin{align}
  \mu_{d,e} - \mu_\text{max} = d + |e| - (g - 1 + \frac{n}{2})
  + \sum_{i = 1}^{n}[(-1)^{e_i} + 1]\alpha_i \ ,
\end{align}
which after substituting $\alpha_i\mapsto -\omega^{(i)}$ becomes
\begin{equation}
\label{eq:mudiffordeven}
\mu_{d,e} - \mu_\text{max} |_{\alpha_i\mapsto -\omega^{(i)}} =d + |e| - (g - 1 + \frac{n}{2}) - \sum_{i = 1}^{n}[(-1)^{e_i} + 1] \omega^{(i)}.
\end{equation}
Coefficients of $\omega^{(i)}$ are all non-negative (actually all $0$) only when $e=(1,1,\cdots,1)$. Similarly, when $n$ is odd, 
\begin{equation}
	\mu_{d, e} - \mu_\text{max} = d + |e| - (g - \frac{3}{2} + \frac{n}{2})
	+ [(-1)^{e_1} - 1] \alpha_1 + \sum_{i = 2}^{n} [(-1)^{e_i} + 1] \alpha_i \ ,
\end{equation}
which become 
\begin{equation}
\label{eq:mudiffordodd}
	\mu_{d, e} - \mu_\text{max}|_{\alpha_i \to \frac{1}{2} + (-1)^{\delta_{i1}} \omega^{(i)}}= d + |e| - (g - \frac{3}{2} + \frac{n}{2})
	+ [(-1)^{e_1} - 1]\left[\frac{1}{2} + \omega^{(1)}\right]
	+ \sum_{i = 2}^{n} [(-1)^{e_i} + 1]\left[\frac{1}{2} - \omega^{(i)}\right] \ ,
\end{equation}
upon the replacement $\alpha_i \to \frac{1}{2} + (-1)^{\delta_{i1}} \omega^{(i)}$. This time, coefficients of $\omega^{(i)}$ are all non-negative only when $e=(0,1,1,\cdots,1)$. Therefore we deduce from conjecture \ref{conj2} that ordinary modules of $V_{g,n}$ correspond to the subset $M^\text{ord}_{g,n}=\{ M_{d,{1,1,\cdots,1}} \}$ (resp. $M^\text{ord}_{g,n}=\{ M_{d,{0,1,1,\cdots,1}}\}$) when $n$ is even (resp. odd), and the conformal weight $h$ of the highest weight state is given by $\mu-\mu_{\text{max}}$ which are $0,\ -1,\ \cdots,\ -\lceil \frac{2g -2 + n}{2} \rceil+1$. Here $\lceil x \rceil$ is the ceiling of $x$. 

Next we study the relation between dimensions of fixed manifolds and modularity of character of ordinary modules. As explained in section \ref{sec:modularmodules}, unrefined characters of ordinary modules of $V_{g,n}$ are solutions of certain MDE, hence also belong to a representation $\bbV^{\text{ord}}_{g,n}$ of the modular group. The dimension  of $\mathbb{V}_{g, n}^{\text{ord}}$ is conjectured in \cite{Beem:2021zvt} to be
\begin{equation}
  \dim \mathbb{V}_{g, n}^{\text{ord}}(\ch_0) = (1 - \delta_{g, 0})g(2g + n - 1) + (1 - \delta_{n, 0}) \lfloor \frac{n - 1}{2} \rfloor (g + n - 1 - \lfloor \frac{n - 1}{2} \rfloor) \ .
\end{equation}
When $g$ and $n$ are small, one can find the MDE explicitly and obtain a basis of $\bbV^{\text{ord}}_{g,n}$ including characters of ordinary modules. When $g$ and $n$ become large, it is difficult to find the explicit expression of the MDE, but one can still get a basis of $\bbV^{\text{ord}}_{g,n}$ by looking at the orbit of $\CI_{g,n}$ under the action of the modular group. One possible basis is given by \cite{Pan:2024bne}
\begin{equation}
	\begin{aligned} \left\{\mathbf{T}_{(2)}^{\ell}S_{(2)}\mathcal{I}_{g,n}\right\}_{\ell=0}^{g}& \ \cup\left\{\mathbf{T}_{(2)}^{\ell}S_{(2)}\mathbf{T}_{(2)}^{g}S_{(2)}\mathcal{I}_{g,n}\right\}_{\ell=0}^{g+2}\cup\left\{\mathbf{T}_{(2)}^{\ell}S_{(2)}\mathbf{T}_{(2)}^{g+2}S_{(2)}\mathbf{T}_{(2)}^{g+1}S_{(2)}\mathcal{I}_{g,n}\right\}_{\ell=0}^{g+4}\\ &\cup\cdots  \cdots\cup\left\{\mathbf{T}_{(2)}^{\ell}S_{(2)}\mathbf{T}_{(2)}^{3g-3+n-2}\cdots S_{(2)}\mathbf{T}_{(2)}^{g+1}S_{(2)}\mathcal{I}_{g,n}\right\}_{\ell=0}^{3g-3+n} \ ,\end{aligned}
\end{equation}
where $\mathbf{T}_{(2)}\equiv e^{-\frac{\pi i}{3}(g-n-1)} T^2 -\mathbf{1}$, $S_{(2)} \coloneqq STS$. We use $T^2$ and $STS$ here because they are present in both $SL(2,\bbZ)$ and $\Gamma^0(2)$. With the given basis, we work out the representation matrices of the generators $T, S$, or $T^2, STS$ of the modular group. The Jordan type of the $T$ (resp. $T^2$) matrix for even (resp. odd) $n$ is
\begin{align}
  [2 + g, 4 + g, ..., 2g - 2 + n + g],& & \text{when} & ~n~\text{is even}\ , \\
  [1 + g, 3 + g, ..., 2g - 2 + n + g],& & \text{when} & ~n~\text{is odd} \ ,
\end{align}
with $\lceil \frac{2g -2 + n}{2} \rceil$ Jordan blocks in total. We note that when $n$ is even, $S$ is diagonalizable (while $T$ has nontrivial Jordan type), but when $n$ is odd, the counterpart $STS$ shares the same nontrivial Jordan type with $T^2$. 

On the Hitchin side,  when $n$ is even,  dimensions of fixed manifolds correspond to ordinary modules are
\begin{equation}
	\dim M_{d, (1, \cdots, 1)} = 0,\ 2,\ \cdots,\ g+n \ ,
\end{equation}
so $\{\dim M_{d,{1,1,\cdots,1}}+g+2\}$ matches the above Jordan type. When $n$ is odd,  dimensions of fixed manifolds correspond to ordinary modules are
\begin{equation}
	\dim M_{d, (0, 1, \cdots, 1)}  = 0, ~ 2, ~ \cdots, ~ g + n -3 \ ,
\end{equation}
so $\{\dim M_{d,{0,1,1,\cdots,1}}+g+1\}$ matches with the above Jordan type as well.  We list a few examples in table \ref{table:g1n1fps-2} and \ref{table:g1n1fps-3} where we see the agreement explicitly.

\begin{table}[h]
\begin{tabular}{c||c|c||c}
$g,n$ & $h$ & $J(T)$ & $\mu-\mu_{max}({\dim})$ \\ \hline
$0,4$ & $0$ & $[2]$ & $0(0)$ \\ \hline
$0,6$ & $0,-1$ &  $[4,2]$ & $-1(2), 0(0)$ \\ \hline
$0,8$ & $0,-1,-2$ &  $[6,4,2]$ & $-2(4), -1(2), 0(0)$ \\ \hline
$1,2$ & $0$ &  $[3]$ & $0(0)$ \\ \hline
$1,4$ & $0,-1$ &  $[5,3]$ & $-1(2),0(0)$ \\ \hline
$1,6$ & $0,-1,-2$ &  $[7,5,3]$ & $-2(4),-1(2),0(0)$ \\ \hline
$2,2$ & $0,-1$ &  $[6,4]$ & $-1(2),0(0)$ \\ \hline
$2,4$ & $0,-1,-2$ &  $[8,6,4]$ & $-2(4),-1(2),0(0)$ \\ \hline
\end{tabular}
\caption{\label{table:g1n1fps-2}Ordinary modules of $V_{g,n}$ with $n$ even. The number in bracket is the dimension of the fixed manifold with the value $\mu-\mu_{max}$.}
\end{table}

\begin{table}[h]
\begin{tabular}{c||c|c||c}
$g,n$ & $h$ & $J(T^2)$ & $\mu-\mu_{max}(\dim)$ \\ \hline
$1,3$ & $0,-1$ &  $[4,2]$ & $-1(2), 0(0)$ \\ \hline
$2,1$ & $0,-1$ &  $[5,3]$ & $-1(2), 0(0)$ \\ \hline
\end{tabular}
\caption{\label{table:g1n1fps-3}Ordinary modules of $V_{g,n}$ with $n$ odd. The number in bracket is the dimension of the fixed manifold with the value $\mu-\mu_{max}$.}
\end{table}

\section{Generalizations to higher rank}
\label{sec:higherrank}

One can also find evidences of the mirror symmetry for higher rank class-$\mathcal{S}$ theories. Notable examples include the Minahan-Nemeschansky $E_6$, $E_7$, $E_8$ theories and our conjecture holds for all these cases\cite{Arakawa_2016}.

In higher rank cases, it is much easier to compute the mixed Hodge polynomials of the character variety. For example, let us consider the $SU(3)$ $\CN=2^\ast$ theory, which is the compactification of $6d$ $A_2$ SCFT on a torus with one minimal puncture. The corresponding VOA $V$ is constructed in \cite{Beem:2013sza, Bonetti:2018fqz, Arakawa:2023cki}. The pure part of its mixed hodge polynomial is
\begin{equation}
q(1+q+q^2).
\end{equation}
Following our conjecture, we predict that there are $3$ simple modules contained in a dimension $6$ representation $\bbV$ of the modular group, and the Jordan type of $T$ matrix is $[3,2,1]$.

On the other side, we observe that all (defect) indices are a linear combination of the following expressions,
\begin{equation}
  \mathcal{I}_{\mathcal{N} = 4\ SU(3)}, \quad
  \frac{\vartheta_4(\mathfrak{b})}{\vartheta_4(3 \mathfrak{b})}, \quad
  \mathcal{I}_W \coloneqq \frac{\vartheta_4(\mathfrak{b})}{\vartheta_4(3 \mathfrak{b})} \left(E_1 \begin{bmatrix}
    -1 \\ b
  \end{bmatrix} + E_1\begin{bmatrix}
    -1 \\ b^2 q^{\frac{1}{2}}
  \end{bmatrix}\right) \ .
\end{equation}
They belong to a six dimensional representation of the modular group $\Gamma^0(2)$, and the Jordan type of $T^2$ is $[3,2,1]$. This provides another example of our conjecture in higher rank. It would be interesting to explore more higher rank examples and establish the mirror symmetry.


\section{Outlook}
\label{sec:outlook}

Our proposal of the 4d mirror symmetry for class-$\mathcal{S}$ theories provides a powerful tool to study the representation theory and modular properties of the corresponding VOA $V_{g,n}$. It would be nice if one can further confirm our prediction, or even proof our conjecture, by using tools like  the free field realization of $V_{g,n}$ proposed in \cite{Beem:2022mde, Beem:2022vfz, Beem:2024fom}. Recently in 3d $\CN=4$ theory, the relation between chiral quantization of Higgs branch and coordinate ring of Coulomb branch is discovered \cite{Costello:2018swh, Beem:2023dub}.  It would be interesting to discuss connections between the 4d mirror symmetry and this 3d ones.

One may also ask what the Coulomb branch counterpart of the log modules are. When $g=0$, each fixed point $M_{d,e}$ is isomorphic $\bbP^{\dim M_{d,e}}$, hence $\dim H^{\ast}(M_{d,e}) = \dim M_{d,e} +1$, and we propose that both simple modules and log modules are bijection to elements of the cohomology ring of fixed points. However, when $g>0$ there the cohomology ring of fixed points are more complicated and the counter part of log modules needs further study.

Another related question is how the modular group appears in the Coulomb branch side. In generalized AD theory, the action of modular group was realized by the automorphism of double affine Hecke algebra acting on the cohomology of fixed points. In class-$\mathcal{S}$ cases, the modular group action may also realized by certain action on the cohomology of fixed manifolds as well. Identifying this action may also help to understand the previous question.

\section*{Acknowledgements}

The authors would like to thank Drazen Adamovic, Tomoyuki Arakawa, Penghui Li, Peng Shan and Dan Xie for helpful discussions. W.Y. is supported by  National Key R\&D Program of China (Grant 2022ZD0117000).
 The work of Y.P. is supported by the National Natural Science Foundation of China (NSFC) under Grant No. 11905301. 

\bibliography{ref}

\end{document}